\documentclass[aps,prc,twocolumn,showpacs,floatfix,nofootinbib,preprintnumbers,superscriptaddress]{revtex4-1}
\usepackage{CJKutf8}
\usepackage{epsfig}
\usepackage{color}
\usepackage{rotating}
\usepackage{natbib}
\usepackage{verbatim}
\usepackage{epstopdf}
\usepackage{graphicx}
\usepackage{changes}
\usepackage{bm}
\usepackage{hyperref}
\usepackage{multirow}
\usepackage{makecell}
\usepackage{enumitem}

\begin{document}

\begin{CJK*}{UTF8}{gbsn}

\title{Bayesian approach to model-based extrapolation of nuclear observables}

\author{L{\'e}o Neufcourt}
\affiliation{Department of Statistics and Probability, Michigan State University, East Lansing, Michigan 48824, USA}
\affiliation{FRIB Laboratory,
Michigan State University, East Lansing, Michigan 48824, USA}

\author{Yuchen Cao (曹宇晨)}
\affiliation{Department of Physics and Astronomy and NSCL Laboratory,
Michigan State University, East Lansing, Michigan 48824, USA}

\author{Witold Nazarewicz}
\affiliation{Department of Physics and Astronomy and FRIB Laboratory,
Michigan State University, East Lansing, Michigan 48824, USA}

\author{Frederi Viens}
\affiliation{Department of Statistics and Probability, Michigan State University, East Lansing, Michigan 48824, USA}

\date{\today}

\begin{abstract}
\begin{description}
\item[Background]
The mass, or binding energy, is the basis property of the atomic nucleus. It determines its stability, and reaction and decay rates. 
Quantifying the nuclear binding is important for understanding the origin of elements in the universe. The astrophysical processes responsible for the nucleosynthesis in stars often take place far from the valley of  stability, where experimental masses are not known. In such cases, missing nuclear information must be provided by theoretical predictions using extreme extrapolations. In order to take full advantage of the  information contained in  mass model residuals, i.e., deviations between  experimental and calculated masses, one can  utilize Bayesian machine learning techniques to improve predictions.

\item[Purpose]
In order to improve the quality of model-based predictions of nuclear properties  of rare isotopes far from stability, we consider the information contained in the  residuals  in the regions where the experimental information exist. As a case in point, we discuss two-neutron separation energies $S_{2n}$ of even-even nuclei.
Through this observable, we assess the predictive power of global mass models towards more unstable neutron-rich nuclei and provide uncertainty quantification of predictions.

\item[Methods]

We consider 10 global models based on nuclear Density Functional Theory with realistic energy density functionals as well as two more phenomenological mass models. The emulators of $S_{2n}$ residuals and credibility intervals (Bayesian confidence intervals) defining theoretical error bars  are constructed using Bayesian Gaussian processes and Bayesian neural networks. We consider a large training dataset pertaining to nuclei whose masses were measured before 2003. For the testing datasets, we considered those exotic nuclei whose masses  have been determined after 2003. By establishing statistical methodology and parameters, we carried out extrapolations towards the $2n$ dripline.

\item[Results]

While both Gaussian  processes and Bayesian neural networks reduce the rms deviation from experiment significantly,
GP offers a better and much more stable performance. The increase in the predictive power of
microscopic models aided by the statistical treatment is quite astonishing:
the resulting rms deviations from experiment on the testing dataset are  similar
 to those of  more phenomenological models. We found that
Bayesian neural networks results are prone to instabilities caused by the large number of parameters in this method. Moreover, since the classical sigmoid activation function used in this approach  has
linear tails that do not vanish, it is poorly suited for a bounded
extrapolation.  The empirical coverage probability curves we obtain match very well the reference values, in a slightly conservative way in most
cases, which is highly desirable to ensure  honesty of uncertainty quantification.
The estimated credibility intervals on predictions make it possible to evaluate predictive power of individual models, and also make quantified predictions using groups of models.

\item[Conclusions]
The proposed robust statistical approach to extrapolation of nuclear model results can be  useful for assessing the impact of current and
future experiments  in the context of model developments. 
The new  Bayesian  capability to evaluate residuals is also  expected
to impact research in the domains where experiments are currently
impossible, for instance  in simulations  of the astrophysical $r$-process.

\end{description}
\end{abstract}


\maketitle
\end{CJK*}

% INTRODUCTION

\section{Introduction}

The knowledge of nuclear binding energy is of fundamental importance for
basic science and applications. Nuclear masses define the extent and details
of the nuclear landscape, determine nuclear stability and decay channels, as
well as decay and reaction rates. Precision measurements of nuclear masses
and moments are needed for tests of fundamental symmetries of nature. In
many cases, however, mass measurements are not practically possible and the
missing information must be provided by theoretical models of nuclear
structure. A good example is in the astrophysical $r$-process that
involve neutron-rich, short-lived nuclei that cannot be accessed
experimentally in the foreseeable future.

Theoretically, there has been noticeable progress in global modeling of
nuclear masses and other nuclear properties. A microscopic approach that is
well suited to providing quantified predictions throughout the nuclear chart
is nuclear Density Functional Theory (DFT) \cite{Ben03}. An effective
interaction in DFT is given by the energy density functional (EDF), whose
coupling constants are adjusted to measured observables \cite%
{Ben03,Erl12a,Goriely2013,Goriely2016,UNEDF2,Afanasjev15,Afanasjev16,Wang2015,Xia17}%
. This global approach can be used to assess the uncertainties on calculated
observables, both statistical and systematic \cite{Dob14,McDonnell2015}.
Such a capability is essential, especially in the context of making
wide-ranging extrapolations into the regions where experiments are
impossible.

A basic test of any EDF parameterization is its ability to reproduce
measured binding energies and other nuclear properties across the nuclear
chart. Typically, the commonly used EDFs yield overall root-mean-square
(rms) deviations between theoretical and experimental masses in the range
from 1 to 5\thinspace MeV \cite{UNEDF2,Afanasjev16}. Currently, the best
overall agreement with experimental masses (rms deviation of 0.56\thinspace
MeV) is obtained with the HFB-31 model of Ref.~\cite{Goriely2016}, which is
rooted in the Skyrme EDF. However, this excellent result has been obtained
at a price of several phenomenological corrections adjusted to the data.

In the context of many structural and astrophysical applications, the
challenge is to carry out reliable model-based extrapolations into the
regions where experimental data are not available. Recently, a critical
assessment of the predictive power of various mass models was performed in
Ref.\thinspace \cite{Sobiczewski2014}. They did not find a correlation
between a model's calibration (measured in terms of a rms mass deviation)
and its ability to predict new masses. Indeed, if a model is merely a
many-parameter formula fitted to experimental data, i.e., it does not have a
sound microscopic foundation, it cannot be expected to provide sound
predictions when it comes to major extrapolations outside the known regions.

In this paper, we investigate the predictive power of current global nuclear
models with respect to two-neutron ($2n$) separation energies $S_{2n}(Z,N)$ of
even-even nuclei. The neutron separation energy is a basic structural
observable that determines the position of the neutron dripline, $r$-process
trajectory, and -- in general -- is a strong indicator of nuclear shell
effects and correlations. Theoretically, $S_{2n}(Z,N)$ is an observable that
is safer to predict than the binding energy. Indeed, in most cases,
deviations between computed and experimental values of $S_{2n}(Z,N)$ are
significantly reduced as compared to mass residuals \cite{UNEDF1}. This
better prediction (lower residuals) in separation energies is to be expected
as many systematic model uncertainties in the binding energy cancel out in
binding energy differences. Here and throughout, we use the term
``residual" to denote the difference
between an experimental value and a model prediction, as can be seen in
definition (\ref{residual}) below. We limit our study to even-even nuclei as
we want to avoid additional complications and uncertainties related to the
choice and treatment of quasi-particle configurations in odd-$A$ and odd-odd
systems \cite{Bonneau2007,Schunck2010,Afanasjev2015a}.

When the experimental two-neutron separation energies $S_{2n}^{\mathrm{exp}%
}(Z,N)$ are known, they can be related to the model predictions $S_{2n}^{%
\mathrm{th}}(Z,N,\vartheta )$ via: 
\begin{equation}
S_{2n}^{\mathrm{exp}}(Z,N)=S_{2n}^{\mathrm{th}}(Z,N,\vartheta )+\delta (Z,N),
\label{residual}
\end{equation}%
where $\vartheta $ is the vector representing all model parameters and $%
\delta (Z,N)$ is the separation-energy residual we were just discussing. The
problem at hand is to calibrate the model (i.e., estimate $\vartheta $),
calculate  the separation energies inside and outside the range of experimental
data, and estimate the uncertainty on predictions.

The residual $\delta (Z,N)$ contains a systematic component due to missing
aspects in the modeling (simplifications, incorrect assumptions, incomplete
physics, etc.) and the statistical component stemming from experimental
errors and uncertainties on model parameters $\vartheta $ resulting from the
model optimization process. In the following, we shall disregard the
experimental errors on separation energies as those are usually well below
theoretical uncertainties.

In many current applications, the statistical uncertainty has been estimated
by means of classical linear regression techniques \cite%
{Dob14,Gao2013,Goriely2014,Kortelainen2015,Niksic15,Haverinen17} or Bayesian
inference methods \cite{McDonnell2015,Schunck2015e}, the latter typically
sharing the same spirit as classical techniques from a modeling standpoint.
We will have more to say in {Section \ref{StatModels}} about how the
Bayesian context distinguishes itself. Systematic errors, seen in the
distribution of residuals as trends, or patterns, are extremely difficult to
estimate as the exact model is not available. The systematic uncertainty on
separation energies is often estimated by an analysis of inter-model
dependences. In the context of nuclear masses, this has been done by
comparing predictions of different DFT frameworks and different EDF
parametrizations \cite{Erl12a,Afanasjev15,Wang2015,Xia17}.

One can improve a theory's predictive power by comparing model predictions
to existing data. Here, a powerful strategy is to estimate residuals by
developing an emulator for $\delta (Z,N)$ using a \textit{training} set of
known masses. An emulator $\delta ^{\mathrm{em}}(Z,N)$ can, for instance, be
constructed by employing Bayesian approaches, such as Gaussian processes,
neural networks, and frequency-domain bootstrap \cite%
{Athanassopoulos2004,Bayram2014,Bayram2017,Yuan2016,Utama16,Utama17,Utama18,Bertsch2017,Zhang2017,Niu2018}%
. The unknown separation energies can then be estimated from Eq.~(\ref%
{residual}) by combing theoretical predictions and estimated residuals: 
\begin{equation}
S_{2n}^{\mathrm{est}}(Z,N)=S_{2n}^{\mathrm{th}}(Z,N,\vartheta )+\delta^{\mathrm{em}}(Z,N).  \label{emulator}
\end{equation}%
It is worth noting that by developing reliable emulators $\delta ^{\mathrm{em%
}}(Z,N)$, which take into account correlations between masses of different
nuclei, one can significantly refine mass predictions and estimate
uncertainties on predicted values \cite{Utama17,Zhang2017,Niu2018}.
Moreover, since the surface of residuals $\delta (Z,N)$ contains important
insights about model deficiencies, by studying the patterns of $\delta ^{%
\mathrm{em}}(Z,N)$ one can make progress in developing higher-fidelity
models.

The paper is organized as follows. Section \ref{Models} lists the mass
models and experimental datasets used in our tests. 
The statistical methodology adopted in our work is outlined in Sec.~\ref{StatMethodology}.
The statistical Gaussian
process and Bayesian neural network frameworks employed to construct $\delta
^{\mathrm{em}}(Z,N)$ are described in Sec.~\ref{StatModels}. The results are
presented in Sec.~\ref{Results}, which also contains the analysis of
advantages and disadvantages of different statistical strategies in the
context of previous work. In particular, Sec.~\ref{Extrapolations} studies
the statistically corrected predictions of the models beyond the range of
available data and discusses their reliability. In Sec.~\ref{Numericals} we
discuss some numerical implementation challenges related to the calibration
of neural networks. Finally, Sec.~\ref{Conclusions} presents conclusions
and perspectives for future studies.

% THEORY

\section{Nuclear Models}
\label{Models}

In the context of meaningful extrapolations in the neutron excess $N-Z$
and/or mass $A$, any underlying theoretical mass model we choose to use should
meet several criteria. First, since such a model is meant to be used in the
regions of the nuclear landscape far from the regions of known masses, it
should be based on controlled many-body formalism employing quantified input
(interaction, energy functional). Second, the theoretical framework should
be capable of reproducing basic nuclear properties  impacting nuclear masses (such as shell structure
and deformations). Finally, the model should be globally applicable
throughout the nuclear chart. For these reasons, we eliminate from our
considerations phenomenological multi-parameter mass formulae, such as \cite%
{Koura2005,Duflo1995} which are directly fitted to experimental data.

The models used and evaluated in this study can be divided into three
groups. The first group contains global mass models FRDM-2012 \cite%
{Moller2012} and HFB-24 \cite{Goriely2013} which are commonly used in
astrophysical nucleosynthesis network simulations. The FRDM is a
representative of well-fitted microscopic-macroscopic mass models. The model HFB-24 is rooted
in a self-consistent mean-field approach with several phenomenological
corrections added. For both models, the rms deviation from experimental
masses is around 0.6\thinspace MeV. As explained in the introduction, this
is about as low as one can expect  without
running into having to make uncomfortably many phenomenological corrections.

The second group contains six microscopic Skyrme-DFT models based on
realistic energy density functionals SkM$^*$ \cite{Bartel1982}, SkP \cite%
{Dob84}, SLy4 \cite{Chabanat1995}, SV-min \cite{Kluepfel2009}, UNEDF0 \cite%
{UNEDF0}, and UNEDF1 \cite{UNEDF1}. For these models, the rms mass deviation
typically ranges from 1.5\,MeV (UNEDF0) to $\sim$5\,MeV (SLy4) \cite{UNEDF1}.

The third group contains four microscopic covariant-DFT models based on
realistic relativistic energy density functionals NL3$^*$ \cite%
{Lalazissis2009}, DD-ME2 \cite{Lalazissis2005}, DD-PC1 \cite{Niksic2008},
DD-ME$\delta$ \cite{RocaMaza2011}. Here, the rms mass deviation varies
between 2 and 3\,MeV \cite{Afanasjev16}. The predictions of the second and third group of models can be found in
the theoretical database MassExplorer \cite{massexplorer}.

The solutions of DFT equations corresponding to localized densities exist only for the neutron chemical potential $\lambda_n<0$  (or $S_{2n}>0$). For $\lambda_n>0$, the nucleus is formally unbound and the standard DFT solution is not meaningful \cite{Dob2013,Michel2008}. The $2n$ dripline,  $S_{2n}=0$, is formally reached for  
$\lambda_n=0$ \cite{Erl12a}.  Consequently, the DFT predictions discussed here are terminated when $\lambda_n$ becomes positive.

%%%%
\begin{figure}[htb]
\includegraphics[width=1.0\linewidth]{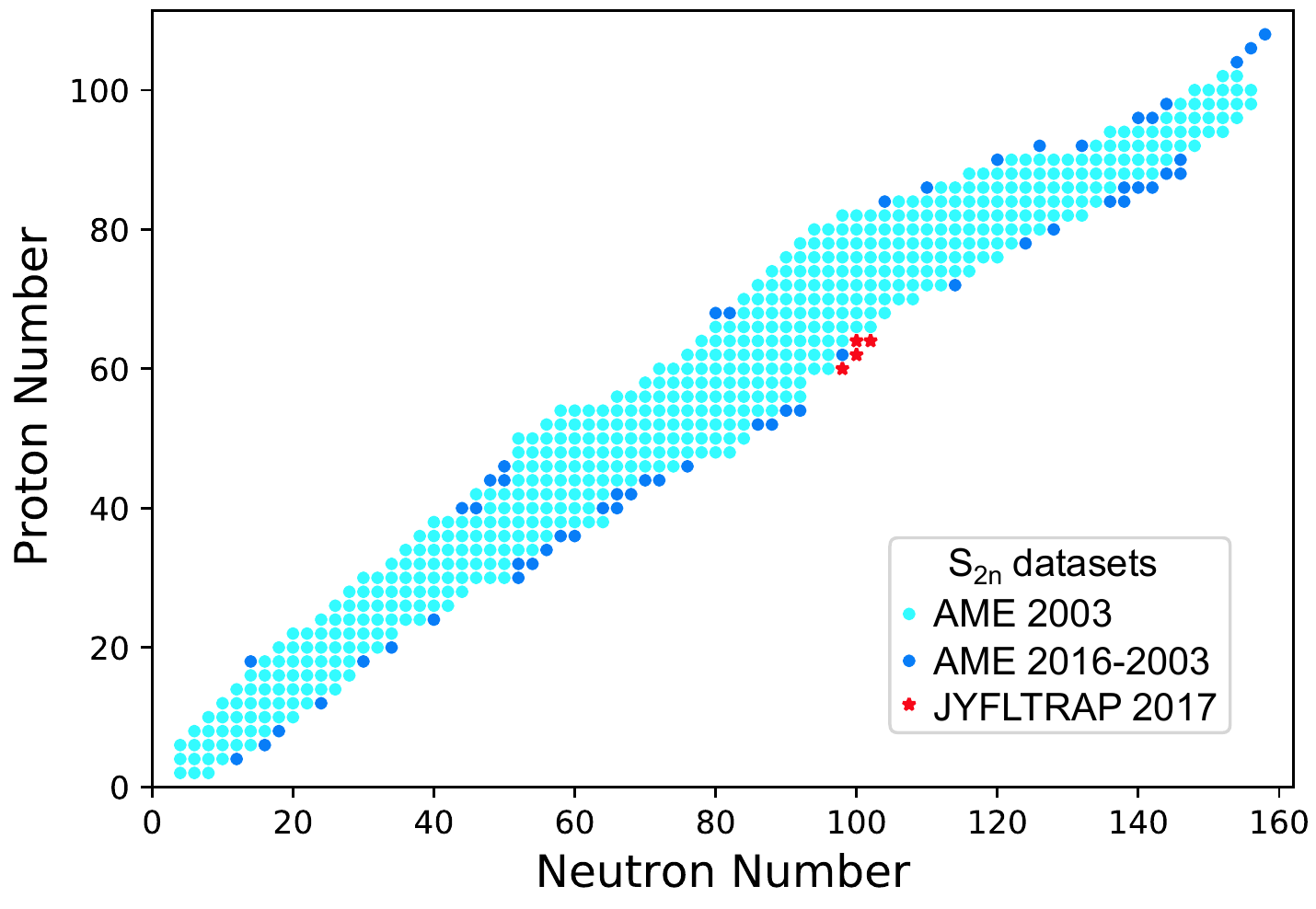}
\caption{The experimental $S_{2n}(Z,N)$ datasets for even-even nuclei used
in our study: AME2003 \protect\cite{AME03a,AME03b} (light dots, 537 points),
additional data that appeared in the AME2016 evaluation \protect\cite%
{AME16a,AME16b} (dark dots, 55 points), and JYFLTRAP \protect\cite{JYFLTRAP}
(stars, 4 points). }
\label{datasets}
\end{figure}
%%%%%%

In order to assess the predictive power of nuclear mass models, for the
training dataset used in Sec.~\ref{AME2003training}, we take 537 values of
$S_{2n}^{\mathrm{exp}}(Z,N)$ from the AME2003 mass evaluation \cite%
{AME03a,AME03b}. As discussed in Ref.~\cite{Sobiczewski2014}, nuclear mass
models based on the mean-field approach perform generally better in heavier
nuclei. Consequently, to test the performance of our statistical models in
heavier nuclei, we also consider the training dataset AME2003-H obtained by
removing the data on lighter nuclei below $Z<20$ from the original AME2003
dataset \cite{Utama16,Utama17,Utama18}. The resulting emulators $\delta ^{%
\mathrm{em}}(Z,N)$ are then used to predict 55 new data points that appeared
in the AME2016 mass evaluation \cite{AME16a,AME16b}, thereby allowing us to
test these predictions as explained in Sec. \ref{AME2003training}. This
joint dataset is referred to as AME2016-AME2003 in the following.
In Sec.~\ref
{AME2016training} we also employed the AME2016-H training dataset ($Z<20$
data removed from AME2016) to test model predictions for 4 points that have
been measured very recently at the JYFLTRAP Penning trap \cite{JYFLTRAP}.

The ranges of experimental data used in this study are shown in 
Fig.~\ref{datasets}. It is to be noted that the mass models employed in this work were optimized to subsets of the masses
listed in the training set AME2003. The only exception is HFB-24, which has also used
additional AME2016-AME2003 data listed in  the evaluation AME2012 \cite{AME12}. 

\begin{figure*}[htb]
\includegraphics[width=0.9\linewidth]{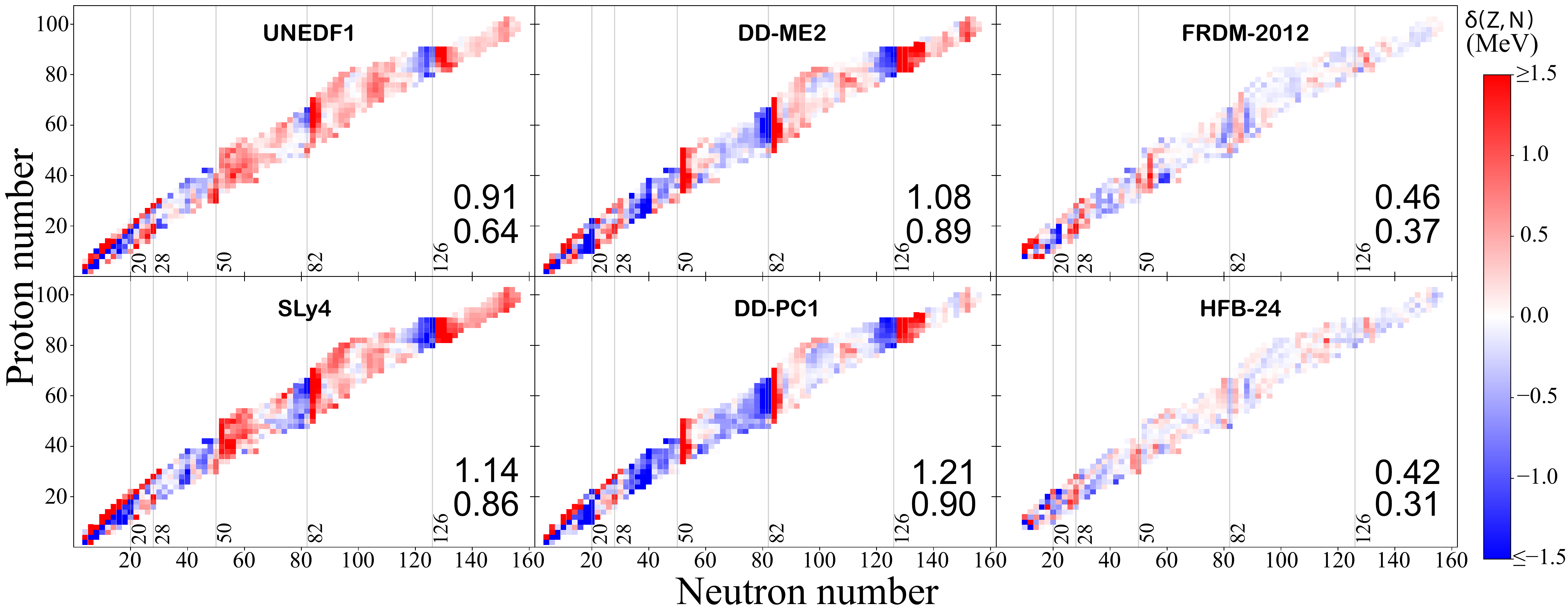}
\caption{Residuals of $S_{2n}(Z,N)$ for the six global mass models with
respect to the testing dataset AME2003. The rms values of $\protect\delta(Z,N)$ in MeV are marked: for AME2003 (upper number) and AME2003-H (lower number)}
\label{raw-residuals}
\end{figure*}

Before moving on to our predictive methodology and results, we describe a
visual inspection of the data. The $S_{2n}(Z,N)$ residuals (\ref{residual})
for the global mass models UNEDF1, SLy4, DD-ME2, DD-PC1, FRDM-2012, and
HFB-24 are shown in Fig.~\ref{raw-residuals}. Long-range patterns are
clearly visible atop the fluctuating background, which can be interpreted as
statistical noise. To better recognize these systematic trends, we smoothed
out the residuals with a Gaussian folding function in $(Z,N)$, see Fig.~\ref%
{smoothed-residuals}. The substantial deviations between experiment and
theory around neutron magic numbers \cite{UNEDF0,Gao2013} are noticeable.
Those are most likely related to an inferior description of the ground-state
collective correlation energies \cite{Bender06,Delaroche2010,Carlsson2013}.
The magnitude and sign of systematic trends in $\delta ^{\mathrm{em}}(Z,N)$
is model-dependent. While the more phenomenological models FRDM-2012 and
HFB-24, primarily fitted to the large mass dataset AME2003, provide a very
good reproduction of experimental values, some long-range deviations are
still present. The models SLy4, DD-ME2, and DD-PC1 exhibit the largest
deviations around the neutron magic gaps, usually attributed to their low
effective masses \cite{Agbemava2014,UNEDF0}. The UNEDF1 model exhibits
fairly smooth systematic trends, with the largest deviations around $N=126$.

%%%%
%%%%%%

%%%%
\begin{figure*}[htb]
\includegraphics[width=0.9\linewidth]{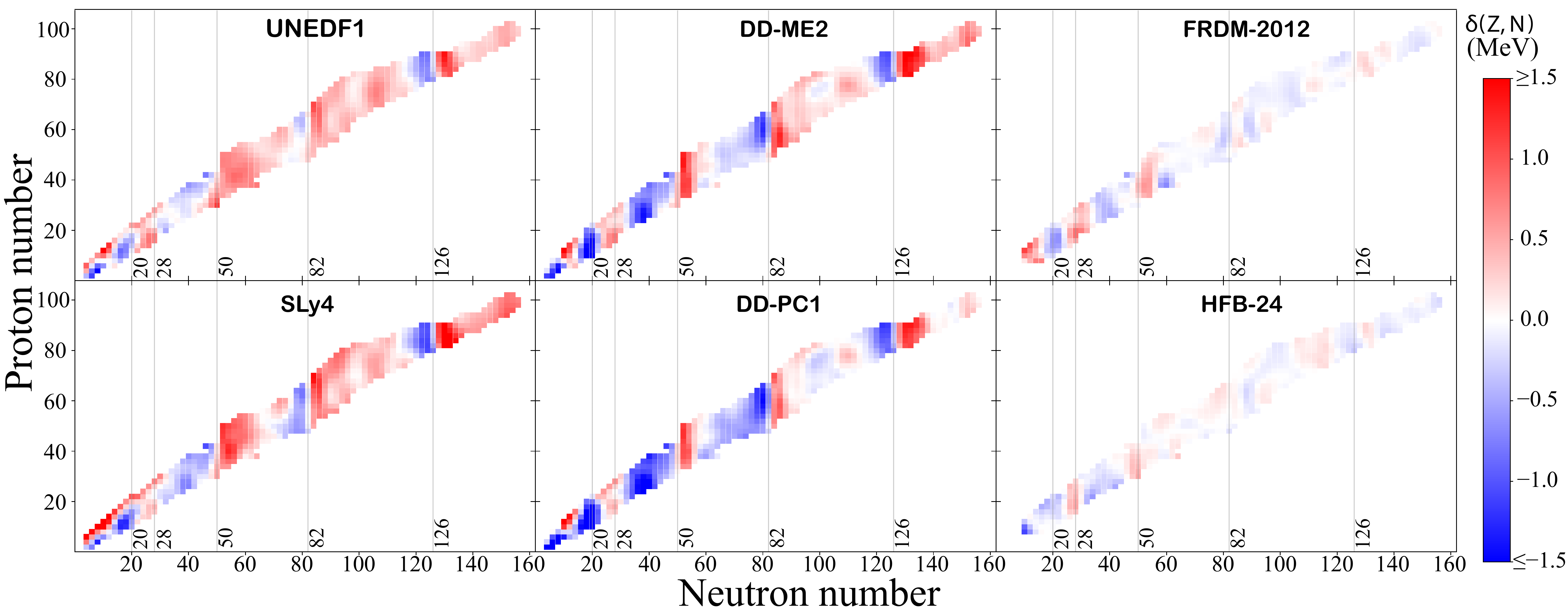}
\caption{Similar as in Fig.~\protect\ref{raw-residuals} but for $S_{2n}(Z,N)$
residuals smoothed with the Gaussian folding function to emphasize
long-range systematic trends. }
\label{smoothed-residuals}
\end{figure*}
%%%%%%

\section{Statistical Methodology}
\label{StatMethodology}

\subsection{Bayesian approach}
\label{StatMethodologyBayes}

A Bayesian methodology can be understood as a statistical solution to a simple
ill-posed inverse problem, when the problem is based on a set of probability
models. In this framework, a model is given which relates observations to
unknown parameters and variables, and contains a term representing
the statistical errors. The model is typically set up as an explanation of the
observations from the unknown parameters and variables, and the question is
to invert the model, i.e., to predict the unknowns from the observations. Since
inverting this equation is an ill-posed problem (too much data, usually far
too much, to determine a unique solution), one must first admit that the
equation contains a noise term, defined in probabilistic terms (using random
variables), in which case one calls it a \emph{likelihood} equation. To
provide a systematical solution which is consistent with the notion of
conditional probability, the Bayesian framework resorts to external information , or beliefs,
about the unknowns: these are probability models, known as the 
\emph{priors}, for each of the unknown parameters and for any unknown variables to be
predicted. The output of the Bayesian analysis is a set of probability
densities or distributions, known as the \emph{posteriors}. In this sense, the 
Bayesian approach provides statistical estimators of all parameters and variables to be
predicted. More precisely, each posterior is a full description of the
uncertainty surrounding each unknown, and in principle, surrounding all the
unknowns jointly (simultaneously) as a group, each posterior mean value
being interpreted as an estimator. Just like in most prediction activities,
the predictive ability of Bayesian approach, based on training data, lies in the
flexibility to first estimate parameters by comparing predicted variables to
known observations, and then use those estimates to make other predictions
in domains where no training data are available. It is important to note that
the Bayesian inversion we just described takes on a particularly simple form, which makes it computationally attractive.

We perform a fully Bayesian analysis of the residuals $\delta (Z,N)$ with
two different classes of statistical models: Gaussian processes (GP), and
Bayesian neural networks (BNN). By  ``fully Bayesian", we mean that we investigate the actual posterior distributions of \emph{all} predicted quantities, and \emph{all}
statistical model parameters. We do not attempt to incorporate any frequentist analysis
within the Bayesian framework, contrary to what is occasionally done, in
particular for GP (see, e.g., Ref.~\cite{KoH}). In this fashion, the full probabilistic
interpretation of the Bayesian output is legitimately preserved. Each of the
two classes of statistical models has its own strength: GP has ability to
take advantage of short-range correlations, and BNN is expected to capture
long-range trends.

Whether GP or BNN, the statistical  model relates the particle numbers $(Z,N)$  to a corresponding observed residual $\delta \left( Z,N\right)$ in the region where data are available to train predictions. Let us fix these ideas by
introducing some generic notation. We will denote each statistical model
class by a function $f$ of the particle numbers  $\left( Z,N\right) $, depending on
parameters $\theta $ which are unknown and must be estimated. Thus, denoting 
$x_{i}:=(Z,N)$ and $y_{i}:=\delta (Z,N)$ for a nucleus label $i$, our
Bayesian model is of the form: 
\begin{equation}
y_{i}=f(x_{i},\theta )+\sigma \epsilon _{i},  \label{eq-model}
\end{equation}%
where $f$ is either GP or BNN with parameters $\theta $, and the
error is modeled as a random variable $\epsilon _{i}$ which is
added to the relation. In general, $f$ can either be a deterministic
function, or a random variable itself. In our GP application described in 
Sec. \ref{GP}, $f$ will be a random variable. For the BNN, $f$ is non-random. We assume that the $\epsilon_{i}$ are independent standard
Gaussian variables (mean zero and unit variance), and $\sigma $ is a noise
scale parameter. The relation (\ref{eq-model}) is called the likelihood equation,
because it relates the data $y_{i}$ with unknown parameters $\theta $
and $\sigma $. We denote the probability density of $y$
in the likelihood model by $p(y|\theta ,\sigma )$, assuming
fixed $\theta $ and $\sigma $. In regions of $x_{i}$ where the values of  $y_{i}$ are
unknown, we can use  (\ref{eq-model})  to predict them. We must also assume prior distributions on the unknown parameters, e.g., a joint probability density 
$\pi (\theta ,\sigma )$.

\emph{Bayes' theorem} states that the posterior density of $(\theta ,\sigma )$
given the data for $y$, under the prior and likelihood models, is
proportional to the product of the likelihood density of $y$ given 
$(\theta,\sigma )$ and the prior density of $(\theta ,\sigma )$:
\begin{equation}
p(\theta ,\sigma |y)\propto p(y|\theta ,\sigma )\pi (\theta ,\sigma ).
\label{eq-bayes}
\end{equation}%
We can also compute predictions $y^{\ast }$ in the regions of  $x$ where the corresponding $y$ is not observed.
This simply requires computing the conditional density of $y^{\ast }$ given 
$y,\theta ,\sigma $, where $y$ is in known regions, and integrating over the
posterior density of the unknown parameters from (\ref{eq-bayes}):
\begin{equation}
p(y^{\ast }|y)=\int p(y^{\ast }|y,\theta ,\sigma )p(\theta ,\sigma
|y) \,d\theta d\sigma .  \label{predict}
\end{equation}%
Typically, and certainly in our case, the conditional probability $p(y^{\ast
}|y,\theta ,\sigma )$ is given explicitly from a direct examination of the
likelihood model (\ref{eq-model}). When an assumption is made in (\ref%
{eq-model}) by which the components of (\ref{eq-model}) are independent of
each other for each  $x_{i}=\left( Z,N\right) $, then 
$p(y^{\ast }|y,\theta ,\sigma )$ does not depend explicitly on $y$. In the GP
case, even though the additive errors terms $\epsilon_{i}$ are independent, the components of $f(x,\theta )$ are not,
because the $(Z,N)$ landscape is viewed as a spatial index
where nearby nuclei must be highly correlated, as we will see in the GP
stochastic specification for $f$. On the other hand, non-explicit dependence
of $p(y^{\ast }|y,\theta ,\sigma )$ on $y$ does hold for the BNN
specification under independence of the $\epsilon _{i}$'s, as we will also
see. However, since the prediction's distribution (\ref{predict}) integrates
out the stochasticity in the parameters' joint posterior density, the final
prediction $p(y^{\ast }|y)$ must depend on $y$, and all is thus in good
heuristic order.

As mentioned earlier, the expressions (\ref{eq-bayes}) and (\ref{predict}) for parameters and predictions are full probabilistic
descriptions. This means that one can compute their mean values, their
credibility intervals, and any other statistics, or any probabilities of
interest. As alluded to in the abstract, a \emph{credibility interval} is the Bayesian counterpart of a confidence interval for classical frequentist statistics; it is an interval in which a parameter or variable has a specified posterior probability of lying. For instance, for each parameter, or each predicted value $y^{\ast}$, one can compute an interval around its Bayesian posterior mean value in which it has a $95\%
$ posterior probability of lying. Such an object, which is akin to a
classical confidence interval, is called a 2-sided $95\%$-credibility interval. 
This credibility level can be replaced by $68\%$, or by a multiple of $%
y^{\ast }$'s standard deviation, or any other level of interest, or
one-sided versions. The same can be done for every parameter. For instance,
if one wonders whether a linear regression or noise scale parameter $\theta
_{1}$ is significantly determined at credibility level $95\%$, one only
needs to check that its one-sided $95\%$ credibility interval excludes the
value $0$. For the sake of conciseness, we do not illustrate these types of
analyses in this paper, though some are implicitly contained in our Bayesian
output. Henceforth, we will use the acronym CI for ``credibility interval''. A reader may think of CI as meaning ``confidence interval'' without running the risk of a major physical misinterpretation.

The prior distribution $\pi (\theta ,\sigma )$ on the model parameters needs
to be thoughtfully designed beforehand according to the physicists'
interpretation and intuition of the model, and any external data, unrelated
to $y_{i}$ as much as possible, which might inform the choice of $\pi$.
This can be a non-trivial exercise. When external data are used to inform 
$\pi$, this results in a hierarchical model. In the present paper, the
absence of such data constrains us to remain in a simple
framework where there is only one level of relationship between data.
However, if one were to push the Bayesian framework further into
the design of nuclear models themselves, then each model designed with
phenomenological information could be considered part of a hierarchical
Bayesian prior. We do not explore this avenue, as it is beyond this paper's
scope. 
Nevertheless our BNN model detailed in \ref{BNNf} is hierarchical in the usual sense, 
since we are considering the neural network weight themselves under their own
posterior distributions, which depend on some (hyper)parameters.
Details are given below on how to choose $\pi $ for each model class
in our simple frameworks.

The main technical challenge in implementing the above basic Bayesian
strategy is to compute posteriors. We rely on a set of Monte Carlo techniques, in which samples from the posterior
distributions are obtained by using 100,000 iterations of an ergodic Markov
chain produced by the Metropolis-Hastings algorithm, an
extension of the Gibbs sampler \cite{Gilks}. Details of the choices, including numerical challenges which
occur in the BNN case, are discussed in Sec.~\ref{Numericals}. Of particular
importance in the implementation is the choice of what to sample first. In
our case, since the Bayesian posterior in Eq.~(\ref{eq-bayes}) is used to 
make predictions in (\ref{predict}), we sample $\theta ,\sigma $ from $p(\theta ,\sigma |y)$ and then we sample $y^{\ast }$ from $p(y^{\ast }|y)$.

\subsection{Predictions and uncertainties}

In the Bayesian methodology, residual corrections and associated uncertainties can be inferred directly
from the posterior samples. In particular, the average value, over all
Monte-Carlo samples, of a specific predicted $y^{\ast }$  provides the correction term that  must be added to 
$S_{2n}^{\mathrm{th}}(Z,N,\vartheta)$ to obtain a prediction for the
two-neutron separation energy at $(Z,N)$. It is
(approximately) equal to the Bayesian posterior mean of interest. Similarly
the sample standard deviation, over all Monte-Carlo samples, of that same 
$y^{\ast}$ gives the one-sigma uncertainty size on the residuals and the corresponding $S_{2n}$ values. This sample standard deviation should be interpreted as what is often referred to as an ``error bar".

Let us be notationally precise. Let $m$ be the total number of Monte-Carlo samples in
our numerical scheme. We denote by $y_{1}^{\ast
}(Z,N),\ldots ,y_{m}^{\ast }(Z,N)$ the ${m}$ posterior Monte-Carlo samples
obtained for  $y^\ast (Z,N)$. Then our prediction for the
residual, for the corrected mass, and for the associated one-sigma
uncertainty level (error bar), are respectively given by:
\begin{eqnarray*}
\delta^\mathrm{em}(Z,N)= &&\frac{1}{m}\sum_{j=1}^{m}y_{j}^{\ast}(Z,N), \\
S_{2n}^{\mathrm{em}}(Z,N)= &&S_{2n}^{\mathrm{th}}(Z,N)+\delta^\mathrm{em}(Z,N), \\
\sigma^\mathrm{em}(Z,N)= &&\sqrt{\frac{1}{m}\sum_{j=1}^{m}\left[y_{j}^\ast
(Z,N)-\delta ^{\mathrm{em}}(Z,N)\right]^2}.
\end{eqnarray*}%

Beyond predictive power, the advantage of our emulator lies in its ability
to quantify uncertainty. In this sense, the ability to compute $\sigma
^{\mathrm{em}}(Z,N)$  belies the far greater power to compute any other
metric for quantifying uncertainty in emulating/predicting $S_{2n}^{\mathrm{em}}(Z,N)$. As mentioned, one can construct a $95\%$ CI for $S_{2n}(Z,N)$ by finding an
interval (for example, centered around the posterior mean $S_{2n}^{\mathrm{em}}(Z,N)$) which contains $95\%$ the corresponding
proportion of all the sampled values $y_{j}^{\ast}(Z,N)$. When the
posterior distribution of $y^{\ast }(Z,N)$ is highly symmetric, this is almost exactly the same as defining the left- and
right-endpoints of the $95\%$-CI for $y^{\ast }(Z,N)$.
When the posterior distribution of $y^{\ast }(Z,N)$ is
very close to normal, one can use the value of $\sigma ^{\mathrm{em}}(Z,N)$ as a shortcut to make the following approximate claims:
\begin{enumerate}[label=(\roman*)]
\item If $y^{\ast }(Z,N)$ is approximately normally distributed, then a
two-sided $95\%$-CI for $S_{2n}(Z,N)$ is approximately 
$S_{2n}^{\mathrm{em}}(Z,N)\pm 1.96\,\sigma ^{\mathrm{em}}(Z,N)$;
\item If $y^{\ast }(Z,N)$ is approximately normally distributed, then a
two-sided $68\%$-CI for $S_{2n}(Z,N)$ is approximately 
$S_{2n}^{\mathrm{em}}(Z,N)\pm\sigma ^{\mathrm{em}}(Z,N)$.
\end{enumerate}
Generally speaking, though it is not common practice given the prevalence of
the above two shortcuts, it is safer to define a $100(1-\alpha )\%$-CI for $S_{2n}(Z,N)$ as an interval containing $\left( 1-\alpha
\right) m$ consecutive samples among the $m$ ordered sampled values $S_{2n}^{\mathrm{th}}(Z,N)+y_{j}^{\ast }(Z,N)$. This avoids miscalculations and
misrepresentations due to assuming that the posterior law of 
$y^{\ast }(Z,N)$ is approximately normal. In our case, we used the shortcut, having checked beforehand that the posterior distributions are approximately normal. 

It is  to be noted that  the quality of a
model  cannot be assessed solely  by  comparing   calculated  expectation values against   known data.
The full prediction  also needs to report its own internal uncertainty quantification (UQ). That UQ is captured by statistics such as $\sigma ^{\mathrm{em}}(Z,N)$, or
CIs as described above. If a $95\%$-CI for $S_{2n}(Z,N)$ contains the experimental value $S_{2n}^{\mathrm{\exp }}(Z,N)$,
that can be taken as a good sign. In the following, we  explain how this
thought can be made more systematic by looking at the systematic predictions
of $S_{2n}(Z,N)$ .

\subsection{Objective and evaluation}\label{Obj}

The above discussion has hinted at how to evaluate a statistical model's
predictive performance. From the nuclear physics perspective, the predictive
power of a model towards more unstable nuclei is a key criterion to assess
its quality. Accordingly, we can use, as a performance criterion, the
improvement on rms error on testing datasets \emph{outside} the training
domain \cite{Utama18}. This is a different strategy than testing samples of interior
points randomly taken out of the training sample - which has been done in
the previous papers \cite{Utama16,Utama17,Bertsch2017,Zhang2017,Niu2018}. In
search of universality, we model the residuals globally on the large domain
of even-even nuclei. This criterion, while appropriate, is similar, in
spirit, to the goal of looking for posterior means, which are as close to
experimental values as possible. But as indicated above, this cannot be the
entire story, because it tends to ignore a methodology's own internal UQ.

Thus, beyond improvements on model predictions, a real understanding of the
residuals requires an honest assessment of a model's ability to accurately
estimate its own uncertainty. Since the Bayesian output can answer any
quantitative question about uncertainty, we can answer this question of 
honesty in UQ in various ways \cite{Gneiting2007,Gneiting2007a}. The simplest and perhaps most intuitively satisfying way to
measure UQ honesty is the following. We compute each model's so-called 
\emph{empirical coverage probability} for a given level $100\ast (1-\alpha )\%$,
e.g., 68\% or 80\% or 95\%, defined as the proportion of the testing data
which actually falls inside CIs with that same 
$100(1-\alpha )\%$ credibility level. If the UQ is honest, that proportion
should be close to the nominal value $100(1-\alpha )\%$. If the
proportion is much lower than the nominal value (e.g., $75\%$ instead of the
nominal $95\%$), this means the UQ is dishonest, because its CIs are far too narrow: they should have been wide enough so that
approximately $95\%$ of the testing data should fall inside the 95\%-CI. The morally charged label of 
``dishonest" is appropriate: using overly narrow CIs implies false claims about a model's high precision, in a
community where every model is ultimately judged by its precision. If the
proportion is much higher than the nominal value (e.g., $99\%$ instead of 
$95\%$), then the UQ is perhaps also dishonest in the sense of being
excessively self-critical: the prediction is not claiming to be more precise
than it really is, but this situation is wasteful because the CIs are wider than they need to be. After all, false modesty is not
necessarily a virtue.

\section{Statistical Models}\label{StatModels}

\subsection{Gaussian Process model\label{GP}}

Gaussian processes have been heavily adopted in recent years in physics and
other natural sciences to model the local structure of complex systems
such as computer models \cite{KoH}. In our context, a GP is a Gaussian
field, i.e., a Gaussian functional on the two-dimensional nuclear domain,
which is characterized by its mean function and covariance function (see
Ref.~\cite{MacKay} for an exhaustive presentation of GP).
GP are designed to capture the spatial structure of the residuals where we
can assume that neighboring nuclei  should have similar properties.
Since the nuclear domain is finite and discrete, 
a GP model is a finite-dimensional Gaussian vector
indexed by particle numbers $(Z,N) $, which distribution is thus given
solely by its mean and covariance matrix. 
We take the mean function to be $0$, and in order to model the ``spatial"
dependence of nearby nuclei in the nuclear landscape, we use an
exponential quadratic covariance kernel 
\begin{equation}
k_{\eta ,\rho }(x,x^{\prime }):=\eta ^{2}e^{-\frac{(Z-Z^{\prime })^{2}}{%
2\rho _{Z}^{2}}-\frac{(N-N^{\prime })^{2}}{2\rho _{N}^{2}}},
\end{equation}%
where $x=(Z,N) $, and the parameters $\theta \equiv \{\eta ,\rho_{Z},\rho _{N}\}$ have a natural interpretation: $\eta $ defines the scale,
i.e. the strength of dependence among neighboring nuclei, and $\rho _{Z}$ and $\rho _{N}$
are  characteristic  correlation ranges  in the proton and neutron direction, respectively.
Note that $k$ is a
\emph{bona fide} covariance matrix because, up to a linear transformation of its
variables, it is the classical Gaussian kernel (also known as Radial Basis
Function or square exponential kernel): the latter has that property because it can be written as
the tensor product of two copies of the function $\exp \left( -\left\Vert
x\right\Vert /2\right) $ multiplied by a sum of {products of monomials which
are symmetric in }$\left( x,x^{\prime }\right) ${. }Other classical families
of kernels include exponential (Laplacian) kernels and Mat\'{e}rn kernels. While all
three families have comparable performance on short range, Mat\'{e}rn kernels involve
Bessel functions resulting in longer computations and Laplacian 
kernels have unneeded heavier tails. Consequently, using the notation 
$\mathcal{GP}$ for a GP as a Gaussian vector, we define the function $f$
 in the model equation (\ref{eq-model}), as the following
random vector with parameters $\theta $ and component index $x=(Z,N)$: 
\begin{equation}
f(x,\theta )\sim \mathcal{GP}(0,k_{\eta ,\rho }),  \label{GPf}
\end{equation}%
which means that the law of the Gaussian vector $f(x,\theta )$ has mean 
$0$ and covariance matrix $k_{\eta ,\rho }$. We emphasize that the
correlation kernel $k_{\eta ,\rho }$ is the key component of the Gaussian
 process $f(x,\theta )$; it is
calibrated on the values $k_{\eta ,\rho }(x_{i},x_{j})$ and used to predict 
$y^{\ast }$ according to the Gaussian conditional distribution of $y^{\ast }$
given $y,\eta ,\rho $ which can be expressed explicitely with $k_{\eta ,\rho}$ \cite{MacKay}. 
Hence in the GP case the noise parameter $\sigma $ in Eq.~(\ref{eq-model}) 
represents the pure experimental uncertainty, which is
negligible with respect to  nuclear model uncertainties
involved, so that it is natural to fix it to the average scale of the
experimental uncertainty (0.0235 MeV). In the case of BNN, as we will see
next, since the model does not have another source of randomness 
the term $\sigma \varepsilon _{i}$ is necessary to account for the
model uncertainty. For GP, model uncertainty is taken into account in the 
$\mathcal{GP}$ term. Moreover, it may not be possible to
extricate the experimental error from the uncertainty of the model if both
were included in GP's specification, since we use only one experimental datum per
nucleus.

\subsection{Bayesian Neural Network}

An artificial neural network (ANN) is a non-linear function $f$ mapping the
data and parameter variables $(x,\theta )$ into a hierarchy of one or more
layers of linear combinations of so-called hidden neurons. In this paper,
because of the relatively limited amount of data (and the prohibitive expense 
associated with the immediate acquisition of
new data), a major consideration is statistical parsimony as a principle to
enhance robustness. Therefore, we limit the number of parameters that need
to be estimated by considering only one hidden layer. Our function
 $f:(x,\theta )\mapsto f(x,\theta )$ with one hidden layer containing $H=30$
hidden neurons \cite{Neal,MacKay} has the following specification: 
\begin{equation}
f(x,\theta ):=a+\sum_{j=1}^{H}b_{j}\phi \left(
c_{j}+\sum_{i}d_{ji}x_{i}\right) ,  \label{BNNf}
\end{equation}%
where $\phi $ is a non-linear activation function (a function whose shape
allows a neuron $j$ to transition abruptly from a low inactivated value to
a high  activated value). The parameters are the linear regression weights 
$\theta \equiv \{a,b_{j},c_{j},d_{ij}\}$, which are both internal to each
neuron $j$'s activation and based on data $x$ and external to allow the neurons to interact as a network. A Bayesian Neural Network (BNN) is simply an ANN with additive
noise, considered as an explanation for the response data $y$, in which the
objective is to compute the posterior distribution of all model parameters (and
unobserved  variables $y$, as a Bayesian prediction). In
other words, as we explained in Sec.~\ref{StatModels}, a BNN is the
Bayesian analysis of the model (\ref{eq-model}) where $f$ is
defined by Eq.~ (\ref{BNNf}).

Recall from Sec.~\ref{StatModels} that one fundamental
difference between GP and BNN is that in GP the specification 
(\ref{eq-model}) contains a stochastic $f$, whereas for BNN $f$ is deterministic.
Furthermore, we assume  that the noise $\epsilon $ in (\ref{eq-model}) for BNN is
a normal vector with independent and identically distributed components,
with zero mean  and unit variances. We presume, with no further
comments beyond invoking the principle of parsimony, that there is no
information gain in BNN in making more complex assumptions about the noise
structure, except to say that such assumptions would take us beyond the
spirit and scope of basic ANN.

Therefore, since the components of $\epsilon $ are independent with unit
variance, the likelihood function is
\begin{equation}
p(y|\theta ,\sigma )\propto e^{-\sum_{i}\frac{(y_{i}-f(x_{i},\theta))^{2}}{%
2\sigma ^{2}}},
\end{equation}
where $\sigma $ is the noise scale in (\ref{eq-model}). In particular, any
two components of $y$ are independent of each other, given $(\theta,\sigma)$. The symbol $y$ in the formula above can be interpreted as the concatenation of what ends up being the
training data $y$ and the predictions $y^\ast$. Therefore, the training
data $y$ and the predictions $y^{\ast }$ are stochastically independent
given $(\theta,\sigma)$. Hence we have $p(y^\ast|y,\theta ,\sigma )=p(y^{\ast }|\theta ,\sigma )$ for BNN as discussed in Sec.~\ref{StatMethodologyBayes}.

As previously noted in Refs.~\cite{Utama16,Utama16bis,Niu2018}, the
accuracy of BNN is much enhanced when the prior weights are given according
to a hyperprior distribution in a Bayesian hierarchical setting (that
hierarchy is not to be confused with what would result from using several
hidden layers in the underlying ANN). Accordingly, we take independent Gamma
prior distributions with unit parameters (hence mean 1) on the weight variances $\gamma_{k}$, centered Gaussian prior
distributions with variance $\gamma_{k}$ on the weights, and another independent Gamma prior
distribution for $\sigma $ with mean 1.

The default BNN typically
assumes a sigmoid activation function $\phi (z)=\tanh (z)$. While the
choice of the activation function has in general a minor impact on a BNN's
performance, the hyperbolic tangent has linear tails which cannot vanish
simultaneously, raising potential issues in the case of a bounded
extrapolation. This is particularly true when one is  modeling the residuals
globally on the large  nuclear domain, in which case it may be more
appropriate to choose a more local activation function, e.g. a Gaussian
kernel function which builds the prediction locally with small bumps that
can capture local trends, similarly to what occurs in a GP.

The number of parameters in a BNN is key to the model's
performance. With about 500 data points, taking $H=30$ neurons leads to much
better performance than higher (or lower) $H$. As  mentioned earlier,
increasing the number of layers beyond $L=1$ decreases performance.
This is almost certainly due to the small amount of data which, as we
explained, is a non-negotiable aspect of this type of nuclear theory UQ
study. The number of parameters for an ANN containing $L$ layers with $H$
hidden neurons in each layer is given by $(1+|x|)H+[H(H+1)]^{L-1}+(H+1)|y|$,
where ${|x|}$ and $|y|$ are the respective dimensions of the network data
input and outputs. With $\left\vert x\right\vert =2$ (or 4 in the
refinement described in Sec.~\ref{refinements}) and $\left\vert {y}\right\vert
=1$,  this results in 121 (or 181) parameters; adding one layer would add
120 parameters at once. There exists an unwritten rule of thumb in
statistics,  by which the ratio of
data to parameters needed in order to have a hope of estimating parameters
in a statistically significant way in linear regressions (e.g., with 95\%
confidence/credibility on most parameters), should be bounded below by 10 in
a classical frequentist setting, and should be bounded below by 3 in a
Bayesian setting when there is no expectation of showing that the output is
insensitive to the priors. With about 500 datapoints, this explains why one
cannot use more than one BNN layer in our study, and why a frequentist ANN
study is impossible. It is also worth noting that the number of parameters
in our GP model is much lower than for BNN,
which immediately provides GP an informal UQ advantage over BNN in our study.

\subsection{Refinements}\label{refinements}

As we will see in Sec.~\ref{Results}, while the basic GP
brings a significantly improved predictive power to nuclear mass
models, the basic BNN performs poorly in terms of noise reduction when it comes to the 
extrapolation problem. Still, several applications have been successful in reducing rms deviations
on masses \cite{Utama16,Utama16bis,Utama17,Utama18,Niu2018}. A major factor
explaining this difference is that Refs.~\cite{Utama16,Utama16bis,Utama17,Niu2018} 
do not measure the prediction error on
extrapolations but rather on a traditional cross-validation subset.

Moreover, these papers (as well as Ref. \cite{Utama18})
systematically disregard light nuclei in both training and testing sets,
resulting in a less global approach: Refs.~\cite{Utama16,Utama16bis,Utama17,Utama18} limit the domain to the isotopic
chains above $^{40}Ca$ while \cite{Niu2018} considers only the nuclei
with $Z$ and $N$ above 8 and experimental errors lower than 100 keV. To
provide comparable results in our framework, we have implemented this data reduction
on both our GP and BNN models, with corrections based on the reduced domain
of nuclei below calcium which we denote by GP(H) and BNN(H).

Additionally Ref.~\cite{Niu2018} has improved the performance of
BNN by enriching the input with information on the nucleus' proximity to magic
gaps. Indeed, as seen in Fig.~\ref{smoothed-residuals}, the largest deviations
between experiment and
theory appear around neutron magic numbers.
Consequently, following Ref.~\cite{Niu2018}, we increase the input
dimension, from two dimensions $(Z,N)$ to four dimensions
by introducing the non-linear transformation $\tilde{x_{i}}\equiv (d_{N}(x_{i}),p(x_{i}))$, where $d_{Z}(x)$ and $d_{N}(x)$ denote the
distance of $x$ to the closest magic proton and neutron number,
respectively. The quantity $p(x)=\frac{d_{Z}(x)d_{N}(x)}{d_{Z}(x)+d_{N}(x)}$
is the promiscuity factor, which is an indicator of collectivity in open-shell nuclei 
\cite{Casten1987}. The resulting variant calculations are respectively
denoted as GP(T) and BNN(T) in the following.
As we will see, those two refinements are determining for BNN
and bring a minor improvement to GP.

%%%%
\begin{figure}[htb]
\includegraphics[width=1.0\linewidth]{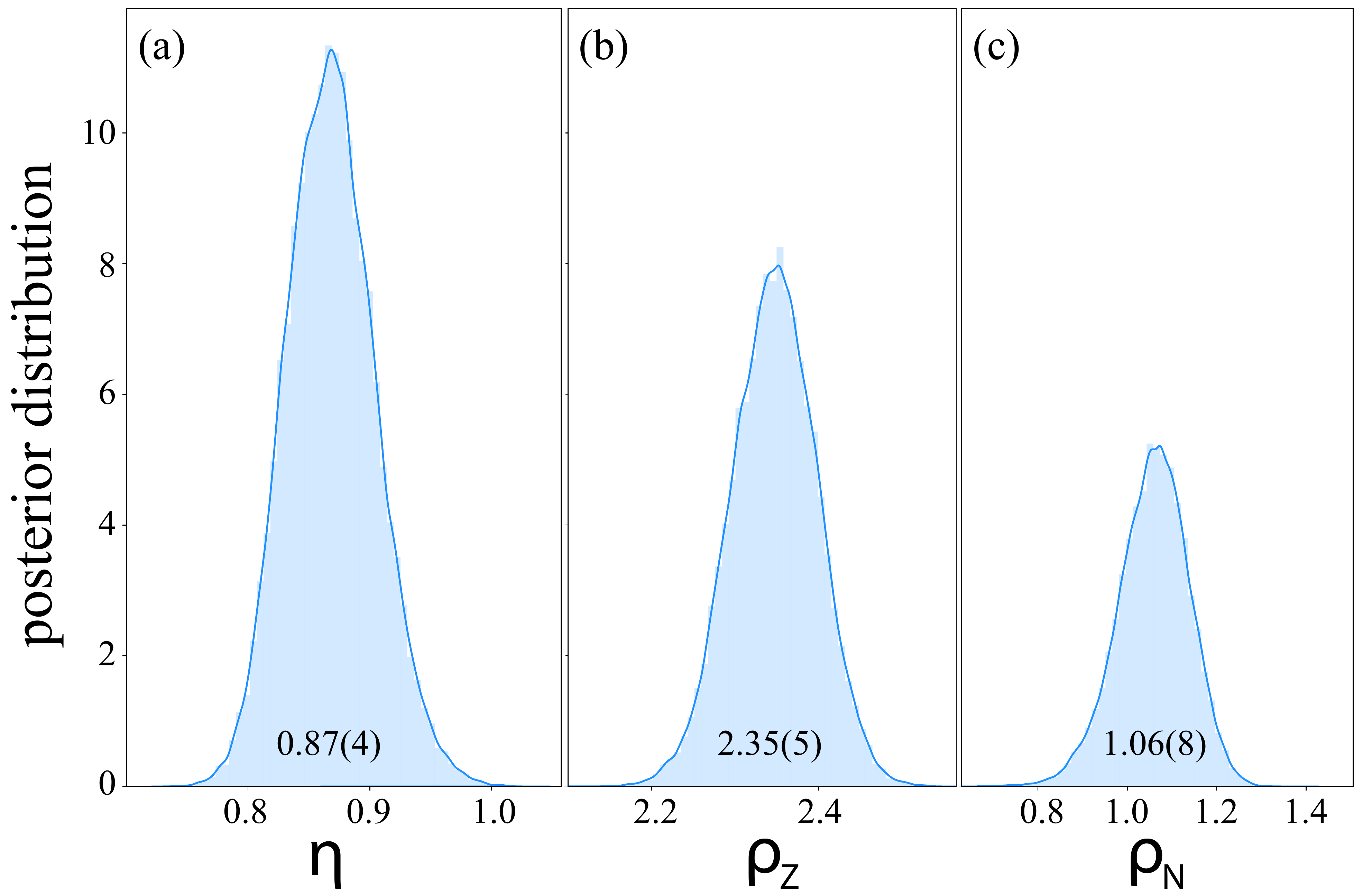}
\caption{Posterior distributions of the GP parameters, with the posterior mean and standard deviation listed.}
\label{GP-posteriors}
\end{figure}
%%%%%%
In Fig.~\ref{GP-posteriors} we show
 the  posterior distributions of the GP parameters in the case of the DD-PC1 model. It is seen that all three parameters are well determined with relatively narrow bell-shaped densities. 
The general scale of the statistical fluctuations is given by $\eta$ at 0.87 MeV. The parameters $\rho_Z$ and $\rho_N$ give the range of the correlation effects  along the $Z$ and $N$ directions, respectively, with precisely $68\%$ concentrated within the neighborhood of size $2\rho$ and $95\%$ within that of size $4\rho$. Here we can assert that about $90\%$ of the correlation effects are located in the region $Z\pm 4$, $N\pm 2$.
We recall that the $\sigma$ parameter in (\ref{eq-model}) scaling the experimental errors was set to 0.0235\,MeV, which is the average of the error bars reported.

%%%%
\begin{figure*}[htb]
\includegraphics[width=0.7\linewidth]{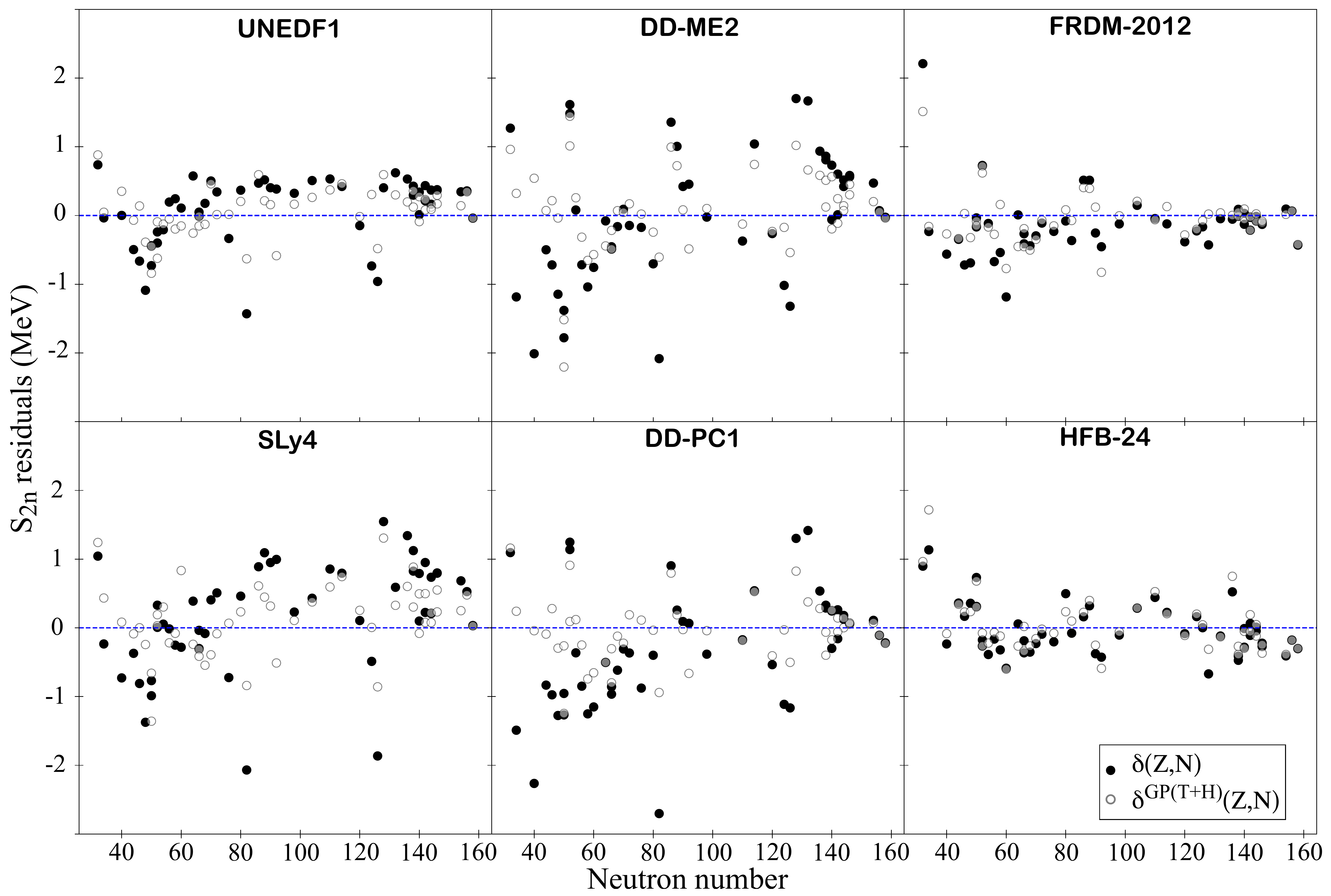}
\caption{Residuals of $S_{2n}(Z,N)$ for the six global mass models with
respect to the testing dataset (AME2016-AME2003): $\protect\delta(Z,N)$
(dots) and the GP emulator $\protect\delta^{\mathrm{GP}}(Z,N)$ (circles). }
\label{GP-residuals}
\end{figure*}
%%%%%%

%%%%
\begin{figure*}[htb]
\includegraphics[width=0.7\linewidth]{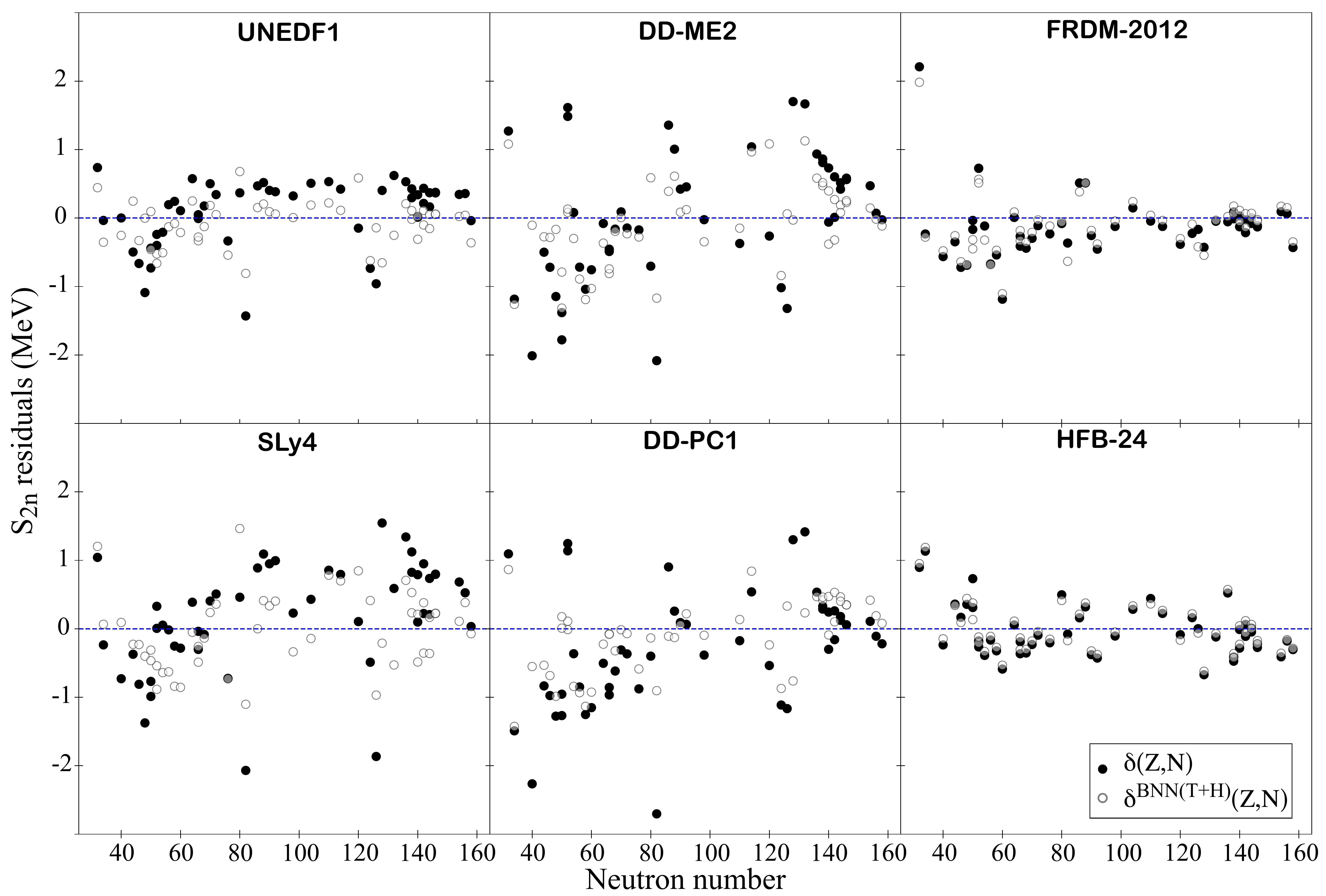}
\caption{Residuals of $S_{2n}(Z,N)$ for the six global mass models with
respect to the testing dataset (AME2016-AME2003): $\protect\delta(Z,N)$
(dots) and the BNN emulator $\protect\delta^{\mathrm{BNN}}(Z,N)$ (circles).}
\label{BNN-residuals}
\end{figure*}
%%%%%%

\section{Results}
\label{Results}

\subsection{Training set: AME2003, testing set: AME2016}
\label{AME2003training}

To test the predictive power of theoretical models and performance of
statistical models, we first carried out calculations involving the training
datasets AME2003 and AME2003-H and the testing dataset AME2016-AME2003. The
results are presented in Figs.~\ref{GP-residuals} and \ref{BNN-residuals} and
Table~\ref{table_2016}.
It is to be noted that AME2016 contains data on remeasured masses since the AME2003 compilation. In some cases,
the differences between old and new data can be significant (up to 30\% difference), especially for light nuclei.
Given the overall consensus that the AME2016 values are more accurate,
the points in question, namely ${}^10$He, ${}^{24}$O, ${}^{34}$Mg and ${}^{52}$Ca, 
are removed from the AME2003 training dataset.
Of course this correction is applied only to the emulator trained on AME2003, 
solely for the purpose of providing meaningful evaluations,
and all other training sets incorporate the most recent measurement available to-date.

Figures~\ref{GP-residuals} and Fig.~\ref{BNN-residuals} show the
residuals of six representative nuclear mass models as a function of the neutron number
before and after statistical corrections with GP(T+H) and BNN(T+H)
respectively. In both GP and BNN, one observes that nearly every
white circle, corresponding to a prediction error of our emulator for a
given nucleus, is closer to the data than its corresponding black dot, representing a global mass model
prediction error without GP or BNN. This indicates that both GP and BNN
corrections improve the predictions systematically. Additionally, several
local trends of the residuals are visibly attenuated, and the distributions
of the corrected residuals are closer to having zero means and to being
independent Gaussian. We observe that the improvement in performance for our
statistical correction is strongest for the relativistic DFT models (DD-ME2
and DD-PC1), and weakest for the more phenomenological models FRDM-2012 and
HFB-24. This is not surprising, as we expect the residuals of more microscopic 
models to exhibit appreciable structure,
whereas there is little hope to improve much the  
phenomenological models which are fitted closely to the data.

Table~\ref{table_2016} lists the rms values of residuals obtained in various
mass models using four different variants, and the two GP and BNN
statistical approaches, for emulators $\delta ^{\mathrm{em}}(Z,N)$. Both GP
and BNN reduce the rms residuals of $S_{2n}$ noticeably, with GP having a
significantly better performance. Both GP and BNN perform best on the
relativistic DFT models  (around 50\% rms reduction), second
on the Skyrme-DFT models (around 30\% rms reduction), and
then on the more phenomenological models FRDM-2012 and HFB-24 which are
already very well optimized to nuclear masses, as we noted 
(around 10\% rms reduction); this is consistent with the corresponding
levels of structure seen in Fig.~\ref{smoothed-residuals}. The increase in
the predictive power of DFT-based models aided by the statistical treatment
of residuals is significant: the rms deviation from experiment in the T+H
variant shows that the carefully optimized Skyrme-DFT models, such as
UNEDF0, UNEDF1, and SV-min, provide results of a similar quality than more
phenomenological models. Overall, for the testing dataset AME2016-2003, the
rms deviation from experimental $S_{2n}$ values is 400-500\,keV in
the GP(T+H) variant for \textit{all} theoretical models employed in this
study, which suggests that our statistical methods capture
most of the residual  structure. At this point, however, we shall re-emphasize the importance of carrying out the full UQ analysis
to assess the quality of a model:  the predicted mean value is certainly not  the whole story.

%%%%%% 
\begin{table}[!htb]
\caption{Root mean square values of $\protect\delta(Z,N)$, $\protect\delta^{%
\mathrm{BNN}}(Z,N)$, and $\protect\delta^{\mathrm{GP}}(Z,N)$ (in MeV) for
various nuclear models with respect to the testing dataset consisting of the
AME2016-AME2003 $S_{2n}$ values. The training AME2003 and AME2003-H datasets
were used to compute the emulators $\protect\delta^{\mathrm{BNN}}(Z,N)$ and $%
\protect\delta^{\mathrm{GP}}(Z,N)$. The two numbers listed under the model's
name in the first column are the uncorrected $\protect\delta_{\mathrm{rms}}$
model values with respect to AME2003 and AME2003-H datasets, respectively.
The rms residuals corrected by a statistical model are shown in the
remaining columns. For each model, GP results $\protect\delta^{\mathrm{GP}}_{%
\mathrm{rms}}$ are given in the upper row and the BNN results $\protect\delta%
^{\mathrm{BNN}}_{\mathrm{rms}}$ are listed in the lower row. The numbers in
parathenses indicate the improvement in percent. The four statistical
variants are listed: Std is the standard treatment with the AME2003 training
dataset; T indicates results involving the non-linear transformation $\tilde{%
x_i}= (d_N(x_i), p(x_i))$; H is based on the reduced dataset AME2003-H
pertaining to heavy nuclei with $Z\ge 20$. }
\label{table_2016}%
\begin{ruledtabular}
\begin{tabular}{c*{4}{l}}
model  & \hspace{7pt}Std  & \hspace{10pt}T  & \hspace{10pt}H & \hspace{3pt}T+H
\\
 \hline \\[-8pt]
\multirow{2}{*}{\makecell{{\bf SkM*} \\ {1.25/1.01}}}
&\small{0.96(23)}	&\small{0.96(23)}	&\small{0.49(52)}	&\small{0.49(52)}\\																		&\small{0.99(20)}	&\small{0.81(35)}	&\small{0.73(28)}	&\small{0.53(47)}\\[4pt]
\multirow{2}{*}{\makecell{{\bf SLy4} \\ 0.95/0.80}}
&\small{0.82(13)}	&\small{0.82(13)}	&\small{0.52(35)}	&\small{0.52(35)}\\
&\small{0.91(3)}	&\small{0.82(14)}	&\small{0.71(11)}	&\small{0.56(30)}\\[4pt]						
\multirow{2}{*}{\makecell{{\bf SkP} \\ 0.84/0.62}}	
&\small{0.75(11)}	&\small{0.75(11)}	&\small{0.38(39)}	&\small{0.38(39)}\\
&\small{0.76(9)}	&\small{0.74(12)}	&\small{0.59(5)}	&\small{0.45(27)}\\[4pt]
\multirow{2}{*}{\makecell{{\bf SV-min} \\ 0.78/0.49}}
&\small{0.70(10)}	&\small{0.70(10)}	&\small{0.32(34)}	&\small{0.33(34)}\\
&\small{0.72(8)}	&\small{1.35(-73)}	&\small{0.50(-1)}	&\small{0.43(12)}\\[4pt]
\multirow{2}{*}{\makecell{{\bf UNEDF0} \\ 0.78/0.54}}	
&\small{0.73(6)}	&\small{0.73(6)}	&\small{0.34(37)}	&\small{0.34(37)}\\
&\small{0.87(-12)}	&\small{0.73(7)}	&\small{0.55(0)}	&\small{0.46(16)}\\[4pt]
\multirow{2}{*}{\makecell{{\bf UNEDF1} \\ 0.66/0.49}}
&\small{0.61(8)}	&\small{0.61(8)}	&\small{0.34(30)}	&\small{0.34(30)}\\
&\small{0.79(-20)}	&\small{0.74(-12)}	&\small{0.53(-10)}	&\small{0.32(33)}\\[4pt]
\multirow{2}{*}{\makecell{{\bf NL3*} \\ 1.19/0.86}}	
&\small{0.84(29)}	&\small{0.84(29)}	&\small{0.46(47)}	&\small{0.45(47)}\\
&\small{1.10(7)}	&\small{0.90(24)}	&\small{0.83(4)}	&\small{0.69(20)}\\[4pt]
\multirow{2}{*}{\makecell{{\bf DD-ME$\bm\delta$} \\ 1.13/0.96}} 	
&\small{0.73(35)}	&\small{0.74(35)}	&\small{0.55(42)}	&\small{0.55(42)}\\
&\small{1.08(4)}	&\small{0.91(19)}	&\small{0.89(7)}	&\small{0.75(22)}\\[4pt]
\multirow{2}{*}{\makecell{{\bf DD-ME2} \\ 1.04/0.95}}	
&\small{0.71(32)}	&\small{0.71(31)}	&\small{0.63(34)}	&\small{0.62(34)}\\
&\small{1.00(4)}	&\small{1.32(-27)}	&\small{0.90(5)}	&\small{0.61(36)}\\[4pt]
\multirow{2}{*}{\makecell{{\bf DD-PC1} \\ 1.10/0.91}}
&\small{0.79(28)}	&\small{0.79(28)}	&\small{0.46(50)}	&\small{0.46(50)}\\
&\small{1.00(9)}	&\small{1.33(-22)}	&\small{0.85(7)}	&\small{0.54(41)}\\[4pt]
\multirow{2}{*}{\makecell{{\bf FRDM-2012} \\ 0.63/0.49}} 	
&\small{0.57(9)}	&\small{0.57(9)}	&\small{0.36(25)}	&\small{0.36(26)}\\
&\small{0.61(4)}	&\small{0.72(-15)}	&\small{0.48(2)}	&\small{0.45(7)}\\[4pt]
\multirow{2}{*}{\makecell{{\bf HFB-24} \\ 0.40/0.37}}
&\small{0.40(-1)}	&\small{0.40(-1)}	&\small{0.40(-8)}	&\small{0.40(-8)}\\
&\small{0.59(-48)}	&\small{0.44(-10)}	&\small{0.37(1)}	&\small{0.35(6)}
\end{tabular}
\end{ruledtabular}
\end{table}
%%%

%%%%
\begin{figure}[tbh]
\includegraphics[width=0.9\linewidth]{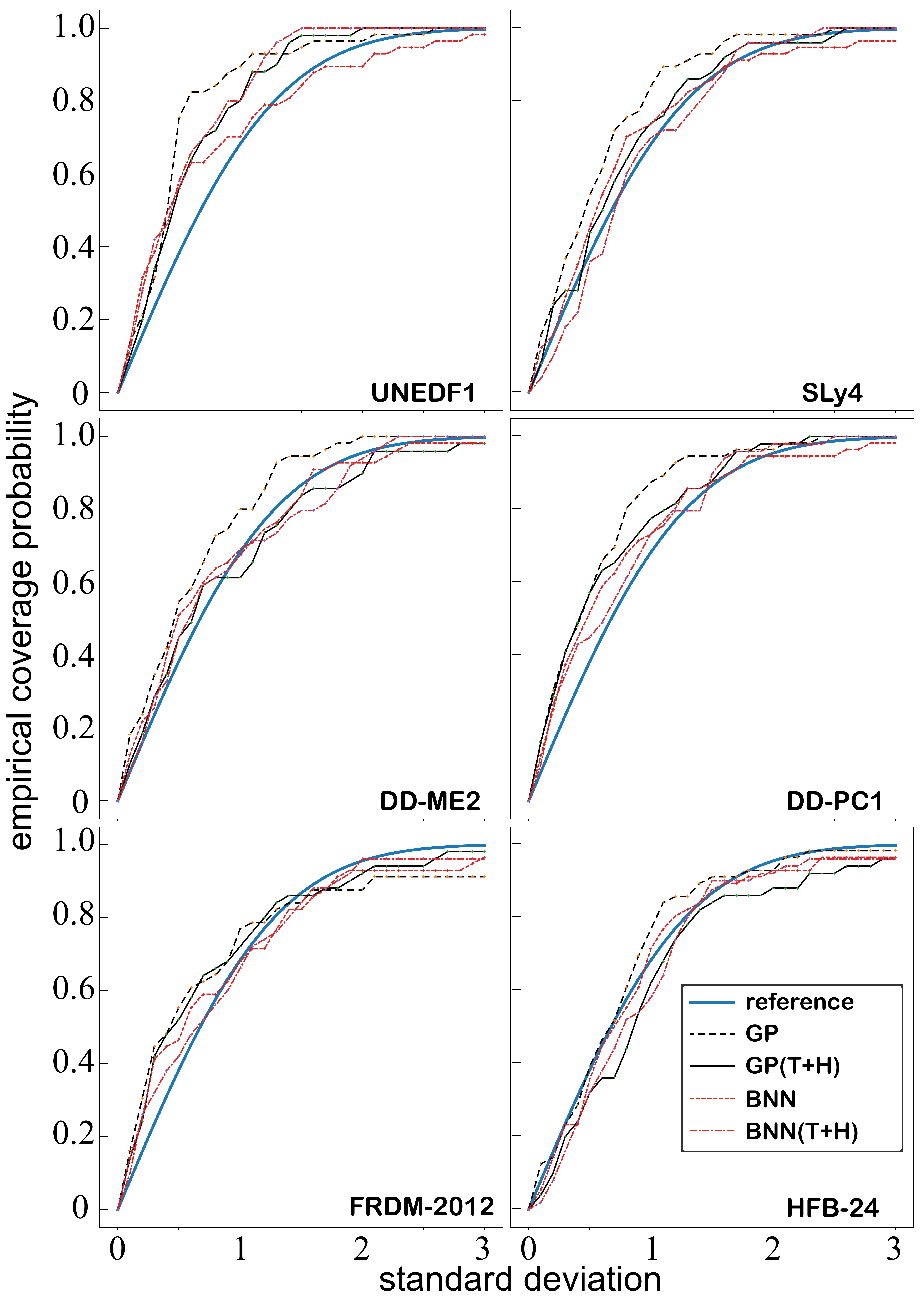}
\caption{Empirical coverage probability for the six models used in our study
as functions of multiples of the standard deviation (s.d.). The reference
curve corresponds to the Gaussian quantiles, i.e., the probability that new
testing data fall into the corresponding CI according to
the domain. That is, at 1\thinspace s.d, the reference curve is 0.68 (and 68
\% of the testing data should fall into the CI); at
2\thinspace s.d. it is 0.95, and so on.}
\label{ecp}
\end{figure}
%%%%%%
Figure~\ref{ecp} shows what is known as empirical coverage probability
(ECP), which is the simple and intuitive metric for assessing the quality of a
statistical model's quantification of uncertainty (see 
Sec.~\ref{Obj} and Refs.~\cite{Gneiting2007,Gneiting2007a}).
 In Fig.~\ref{ecp}, for every model, the  
reference curve shows the fraction of predictions which
should theoretically fall in a CI centered around the
posterior mean prediction, for any given interval width (measured in
posterior standard deviations under normal distribution). In that figure,
every one of the other four curves shows the actual fraction of the
residuals of the testing data that belong to each such CI
for BNN and GP, respectively, with and without the T+H variant. A prediction
point above the reference curve represents a posterior CI
which is too wide because it covers too many points. Thus, it can be
considered as a prediction which is too conservative (or too pessimistic).
As we mentioned before, while not necessarily dishonest, this could be
considered potentially wasteful. A point below the reference curve represent
a CI which is too narrow, which is too liberal (or too
optimistic). This should be
considered dishonest, since it is claiming a level of assurance which is
higher than it should be. Values for the empirical proportions, which
are close to the nominal values of the reference curve are desirable.
In fact, to guard against the risk of giving predictions which are slightly
too optimistic, one is better off hoping for ECPs which are slightly
conservative. At the level of discussing uncertainty on the uncertainty,
aiming for slightly conservative CIs
increases the chances that one's predictions are
sufficiently honest and not very wasteful. 

This objective is, in fact, quite what we observe in Fig.~\ref{ecp}. Regardless of the nuclear physics model or statistical method considered, the distribution of the testing data matches closely the
CIs predicted. The predicted CIs are
slightly conservative for most models - the empirical curve is slightly
above the reference curve - with the exception of HFB-24. Indeed, since HFB-24 matches closely the training data due to its
fairly phenomenological nature, the statistical uncertainty estimated is
very low and does not represent accurately the uncertainty on points which
have not been used in the fit. Overall, the shape of the ECP curves clearly
validates the honesty of our approach, and supports using our corrected
predictions for future measurements with Bayesian CIs.

\subsection{Training set AME2016-H, testing set JYFLTRAP data}
\label{AME2016training}

%%%%%% 
\begin{table}[!htb]
\caption{Similar as in Table~\protect\ref{table_2016} except for the rms values of $\delta(Z,N)$, $\delta^{\mathrm{BNN}}(Z,N)$, and $\delta^{\mathrm{GP}}(Z,N)$ (in MeV) for various
nuclear models with respect to the testing dataset consisting of the four
JYFLTRAP $S_{2n}$ values. The second column shows the uncorrected rms value $\delta_{\mathrm{rms}}$. For each model, the training datasets
AME2003-H (third column) and AME2016-H (fourth column) were used to compute $\delta^{\mathrm{GP}}_{\mathrm{rms}}$ (upper row) and $\delta^{\mathrm{BNN}}_{\mathrm{rms}}$ (lower row) using the T+H variant of
statistical calculations. }
\label{table_2017}%
\begin{ruledtabular}
\begin{tabular}{ccll}
model  & $\delta_{\rm rms}$  & \hspace{1pt}2003-H   & \hspace{1pt}2016-H 
\\
 \hline \\[-8pt]
 \multirow{2}{*}{\textbf{SkM*}} 	 		&\multirow{2}{*}{0.91}	&\small{0.40(56)}	&\small{0.31(66)}\\															
								&					&\small{0.24(74)}	&\small{0.25(72)}\\[3pt]
 \multirow{2}{*}{\textbf{SLy4}} 	 		&\multirow{2}{*}{0.27}	&\small{0.09(65)}	&\small{0.09(67)}\\															
								&					&\small{0.42(-57)}	&\small{0.28(-4)}\\[3pt]
 \multirow{2}{*}{\textbf{SkP}}  			&\multirow{2}{*}{0.19}	&\small{0.16(14)}	&\small{0.14(26)}\\															
								&					&\small{0.35(-85)}	&\small{0.36(-92)}\\[3pt]
 \multirow{2}{*}{\textbf{SV-min}}  		&\multirow{2}{*}{0.14}	&\small{0.11(18)}	&\small{0.10(29)}\\															
								&					&\small{0.17(-20)}	&\small{0.26(-86)}\\[3pt]
 \multirow{2}{*}{\textbf{UNEDF0}}  		&\multirow{2}{*}{0.11}	&\small{0.11(-3)}	&\small{0.11(1)}\\															
								&					&\small{0.33(-199)}	&\small{0.22(-97)}\\[3pt]
 \multirow{2}{*}{\textbf{UNEDF1}} 		&\multirow{2}{*}{0.26}	&\small{0.17(36)}	&\small{0.14(48)}\\															
								&					&\small{0.09(64)}	&\small{0.13(50)}\\[3pt]
 \multirow{2}{*}{\textbf{NL3*}}  			&\multirow{2}{*}{0.32}	&\small{0.19(39)}	&\small{0.22(32)}\\															
								&					&\small{0.17(47)}	&\small{0.18(43)}\\[3pt]
 \multirow{2}{*}{\textbf{DD-ME$\bm\delta$}}  	&\multirow{2}{*}{0.16}	&\small{0.08(50)}	&\small{0.09(46)}\\														
								&					&\small{0.18(-14)}	&\small{0.28(-4)}\\[3pt]
 \multirow{2}{*}{\textbf{DD-ME2}} 		&\multirow{2}{*}{0.30}	&\small{0.12(58)}	&\small{0.13(55)}\\															
								&					&\small{0.28(8)}	&\small{0.29(2)}\\[3pt]
 \multirow{2}{*}{\textbf{DD-PC1}}  		&\multirow{2}{*}{0.28}	&\small{0.17(41)}	&\small{0.13(52)}\\														
								&					&\small{0.25(12)}	&\small{0.27(5)}\\[3pt]
 \multirow{2}{*}{\textbf{FRDM-2012}}		&\multirow{2}{*}{0.13}	&\small{0.10(20)}	&\small{0.09(26)}\\															
								&					&\small{0.05(60)}	&\small{0.05(58)}\\[3pt]
 \multirow{2}{*}{\textbf{HFB-24}}		&\multirow{2}{*}{0.13}	&\small{0.12(2)}	&\small{0.11(12)}\\															
								&					&\small{0.07(43)}	&\small{0.10(25)}\\[3pt]
\end{tabular}
\end{ruledtabular}
\end{table}
%%%

We now investigate the impact of the extended training dataset on an
extrapolation outcome. To this end, we compare predictions based on
AME2003-H and AME2016-H training datasets on the recently measured masses at 
JYFLTRAP. The results are summarized in Table~\ref{table_2017}. The model
rms residuals $\delta _{\mathrm{rms}}$ follow the trend discussed in Sec.~\ref{AME2003training}. Namely, the models FRDM-2012 and HFB-24 make
predictions very close to the data ($\delta _{\mathrm{rms}}=0.13$\,MeV) as well as the recently developed EDFs UNEDF0, SV-min, and DD-ME $\delta $ ($\delta _{\mathrm{rms}}=0.11-0.16$\,MeV). Overall, the
GP approach reduces the rms residuals significantly. This is consistent with Fig.~\ref{raw-residuals}, which shows that the local surface of $\delta (Z,N)$ is
fairly smooth in the region of JYFLTRAP data ($Z\sim 62, N\sim 100$). 
On other other hand, the BNN method is not effective: for  SLy4, SkP, SV-min, UNEDF0, and DD-ME$\delta$ one can see a deterioration of results. This is indicative of a sensitivity of BNN to long-scale correlations that can result in numerical instabilities discussed in Sec.~\ref{Numericals}.

The difference between results based on AME2003-H and AME2016-H training
datasets is insignificant. Overall, we find that the more recent mass
measurements contained in the AME2016-AME2003 dataset do not impose
constraints that are strong enough to modify predictions in the a smooth region of the mass surface. A similar
conclusion was reached in Ref.~\cite{McDonnell2015} in the context of
Bayesian model studies.

\subsection{Extrapolations}
\label{Extrapolations}

As a  follow-up to the two previous exercises we now
train the statistical emulators on the full set of available data for heavy nuclei, i.e., 
AME2016-H  and JYFLTRAP. The
simulations described in Secs.~\ref{AME2003training} and \ref{AME2016training}, on the testing sets, serve as a validation that our
methodology is sound from a UQ perspective and is capable of providing  accurate
predictions.
Perhaps the most important element of our  UQ  is the reliability
of our CIs,
which was assessed in Sec.~\ref{AME2003training} by the analysis of
ECPs at all credibility levels. 
Since our UQ is only slightly conservative,  it essentially preserves
the methodology's full predictive (extrapolation) power.

The question at hand is how far can one extrapolate to provide reliable predictions. 
One can adopt several
approaches to answer this question according to the particular problem of
interest, but the general foundation is that we should trust the obtained Bayesian
CIs  since they have been  validated by our analysis of
the ECPs on points outside the training domain
(see Sec.~\ref{AME2003training} and Fig.~\ref{ecp}), 
with the limitation that the testing points were relatively close to the
training domain.

%%%%
\begin{figure*}[htb]
\includegraphics[width=0.8\linewidth]{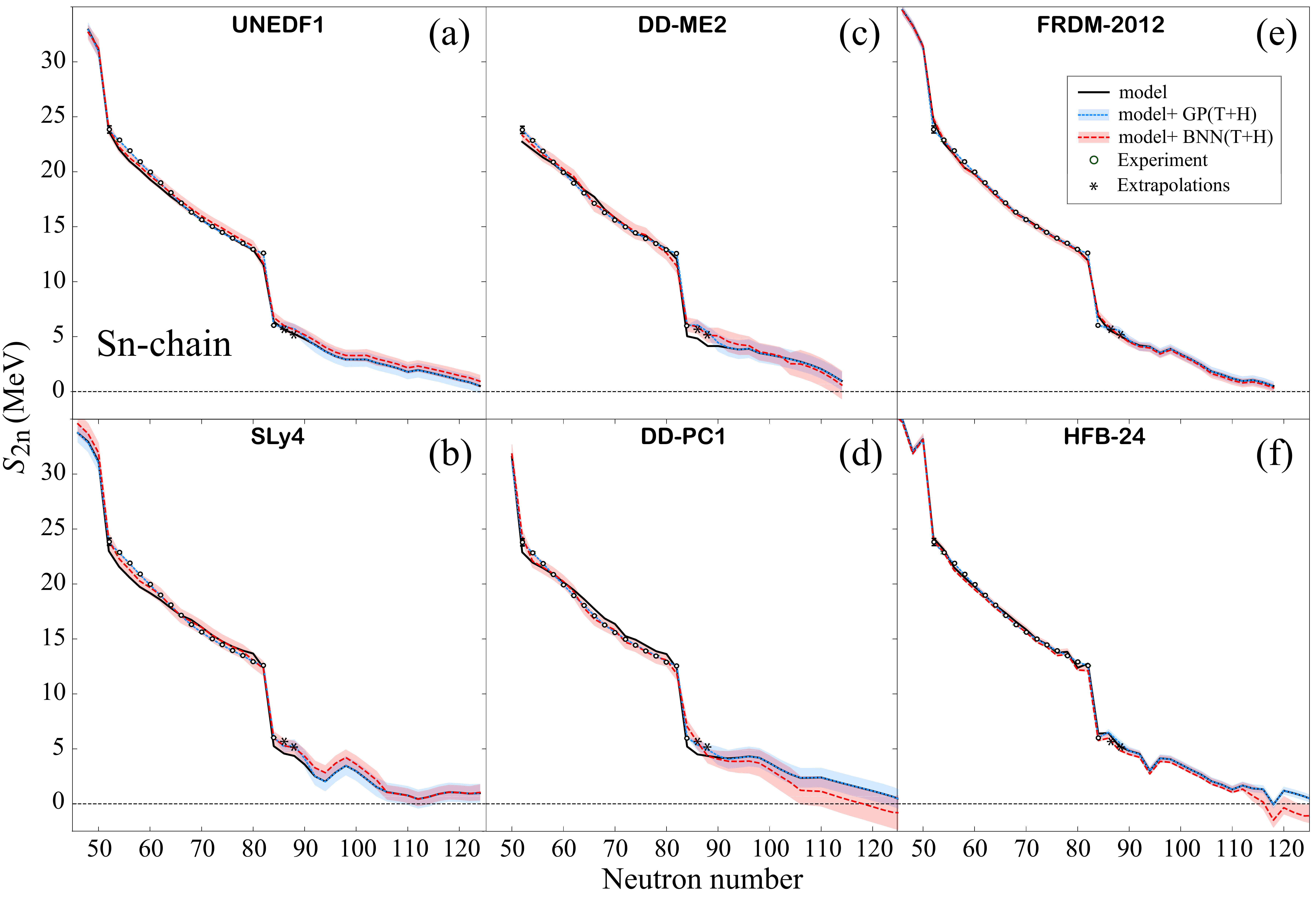}
\caption{Extrapolations of $S_{2n}$ for the even-even Sn chain
calculated  with the six global mass models  with statistical correction $\delta^{\mathrm{em}}$ and one-sigma CIs obtained with GP(T+H) and BNN(T+H).
Experimental (circles) and extrapolated (asterisks)  values from AME2016 \cite{AME16b} are marked.}
\label{extrapolations-sn}
\end{figure*}
%%%%%%

Figure~\ref{extrapolations-sn}
shows the extrapolative predictions of GP and BNN for the six
representative global mass models  for the Sn chain.
As discussed in Sec.~\ref{Models}, the DFT calculations are terminated when $\lambda_n$ becomes positive. This provides a rigid cutoff on mass model predictions.
The Bayesian models estimate  both corrections $\delta^{\mathrm{em}}$ to mass model results as well as  error bars expressed in terms of CIs.
Since  the GP model is fairly local, by construction it goes faithfully  through experimental points and the corresponding  value $\delta^{\mathrm{em}}$
vanishes shortly outside the experimental data range. 
This is not the case for the BNN approach, which is more sensitive to  long-range trends. Still, GP and BNN results are fairly close, and their CIs overlap in most cases. 
The empirical values for $^{136,138}$Sn obtained from extrapolations in Ref.~\cite{AME16b} are very well reproduced  by both GP and BNN.
The size of CIs is consistent with
the pattern of ECPs in Fig.~\ref{ecp}: the largest error bars are obtained for the relativistic DFT models (DD-ME2
and DD-PC1), and smallest for the more phenomenological models FRDM-2012 and
HFB-24. The error bars increase steadily when going away from the experimentally known region.

The small CIs predicted for
HFB-24 require a comment. As seen  in Fig.~\ref{ecp},  HFB-24 is the only model to slightly
underestimate the uncertainties. In fact, HFB-24 has been  fitted to the AME2012 dataset that matches closely the
training dataset AME2003. This causes our
emulator to be blind to the actual underlying uncertainty of HFB-24 outside
its training domain and results  both in  an underevaluation of the
uncertainties on AME2016-AME2003 visible in Fig.~\ref{ecp} and the illusion
of smaller CIs on the extrapolations in Fig.~\ref{extrapolations-sn}. In the context of a discussion about UQ, one would be
tempted to reject the use of HFB-24 in the context of the statistical analysis because it is not honest enough;  to avoid doing so, one would want to incorporate an additional error term in
the statistical model based on HFB-24, one which takes into account
additional uncertainty when making predictions outside of the training
domain. In general, for highly parameterized  models that are very well fitted to experimental data, the statistical approach described in this paper is not going to improve much as  the 
random term in Eq.~(\ref{eq-model}) becomes comparable with the function $f$ describing systematic patterns of model residuals.

A direct  inspection of the CIs in  Fig.~\ref{extrapolations-sn} shows the most conservative
estimate of the location of the $2n$-dripline ($S_{2n}=0$) around $N=104$, as in BNN  with SLy4 and 
DD-PC1. If one were to stick to the GP approach, one would place the $2n$-dripline around $N=120$. The flatness of the posterior mean curves for our statistical emulators
for large neutron numbers
implies that any quantified determination of the location of the $2n$ dripline will be
rather highly sensitive to the size of that uncertainty (e.g., the posterior standard deviation). 
This flatness also implies that one should
decide whether the one-sigma intervals
are at a sufficiently high level of credibility. One-sigma error bar implies 68\% chance  that the true value of predicted  quantity  falls within estimated error bars.
 One might consider using 
two-sigma intervals, corresponding to a CI
of roughly 95\%. The
flatness of the prediction curve would then significantly decrease the drip
line location.

%%%%
\begin{figure}[htb]
\includegraphics[width=0.8\linewidth]{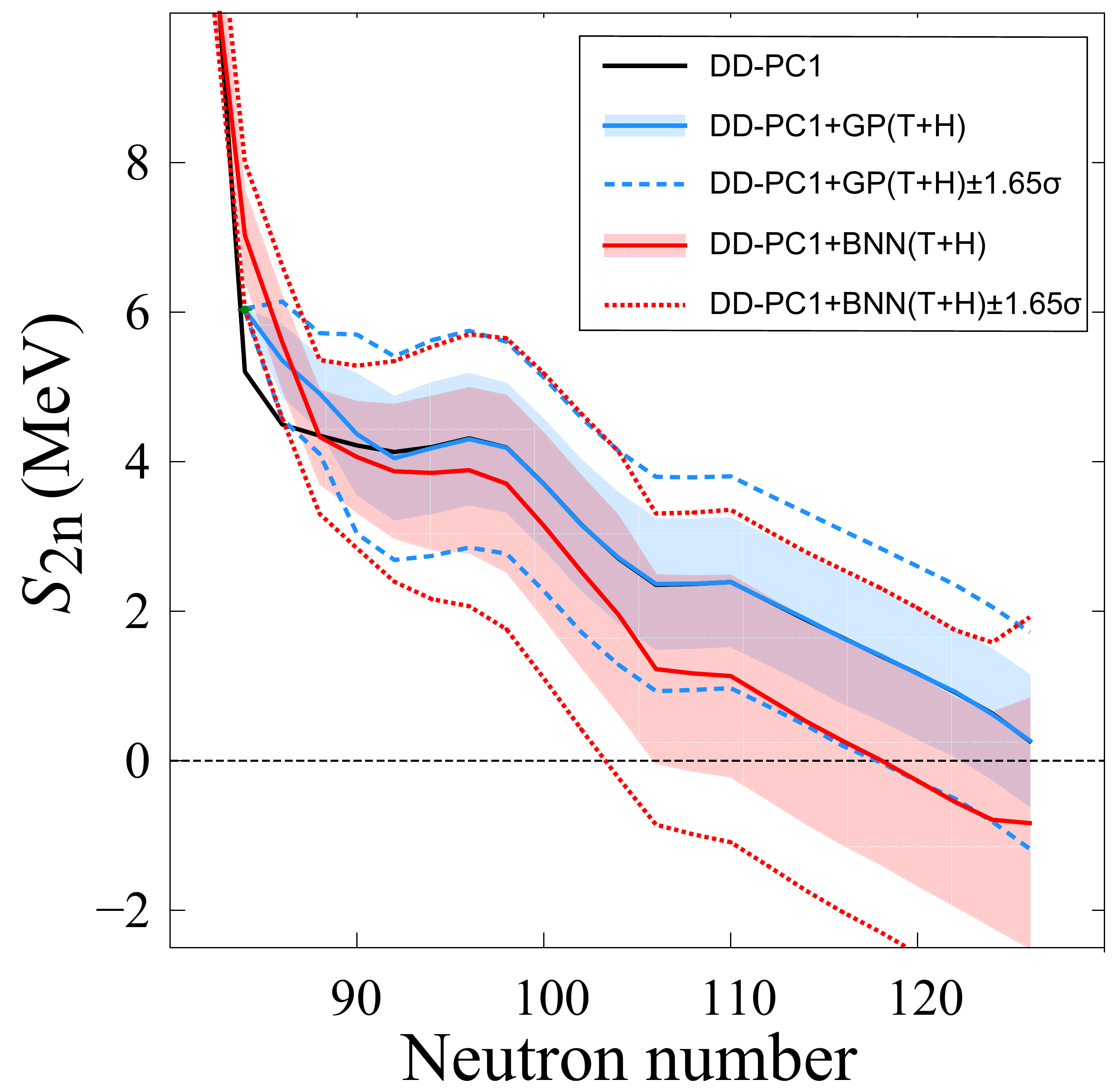}
\caption{Extrapolations of $S_{2n}$ for the even-even Sn chain
calculated  with DD-PC1  with statistical GP(T+H) and BNN(T+H) approaches.
One-sigma and 1.65-sigma CIs are marked.}
\label{driplineDDPC1-sn}
\end{figure}
%%%%%%
If the  objective  is to predict the location of $2n$ dripline for an isotopic chain, a one-sided CI may be more appropriate.
For fixed $Z$, the task is to find the
largest value $N^*$such that the Bayesian posterior probability that
the dripline is below $N^*$ exceeds $1-\alpha $.
The answer to this question would then be roughly equivalent to finding the
value $N^*$ such that the endpoint of its one-sided 
$1.65$-sigma CI barely touches the $S_{2n}=0$ line. This
procures a predictive advantage over simply reading measurements off of
two-sided CIs, since, by extending a one-sided interval to
the level 1.65-sigma, one reaches a credibility of $95\%$, i.e., odds of about 20 to 1 for being right about a dripline. Figure~\ref{driplineDDPC1-sn}
illustrates this approach for the dripline of Sn isotopes predicted with DD-PC1 using 
the statistical GP(T+H) and BNN(T+H) methods. In this case, the DD-PC1 model  and DD-PC1+GP(T+H) predict the $2n$ dripline at $N^*=126$ for its posterior mean value (with $N^*=122$ and  118 at the 1-sigma and 1.65-sigma levels, respectively) while the  DD-PC1+BNN(T+H) model gives a prediction of $N^*=118$
($N^*=102$ and  104 at 1-sigma and 1.65-sigma level, respectively). This discussion demonstrates the na\"{\i}vety of  the absolute declarations, such as: ``DD-PC1 predicts the $2n$ dripline at $N=126$." Note that Figure~\ref{driplineDDPC1-sn} also contains the right-hand limit of the 1-sigma and 1.65-sigma CIs, but this information is not needed to interpret a lower credibility limit on a dripline. Again, as stated, to be safe, we recommend using the neutron-number location where the 1.65-sigma CI crosses the zero-$S_{2n}$ level as a lower limit for the dripline, since the probability of a model predicting the dripline being at this location or below is at least $95\%$.

\subsection{Numerical considerations}
\label{Numericals}

\begin{figure}[htb]
\includegraphics[width=0.7\linewidth]{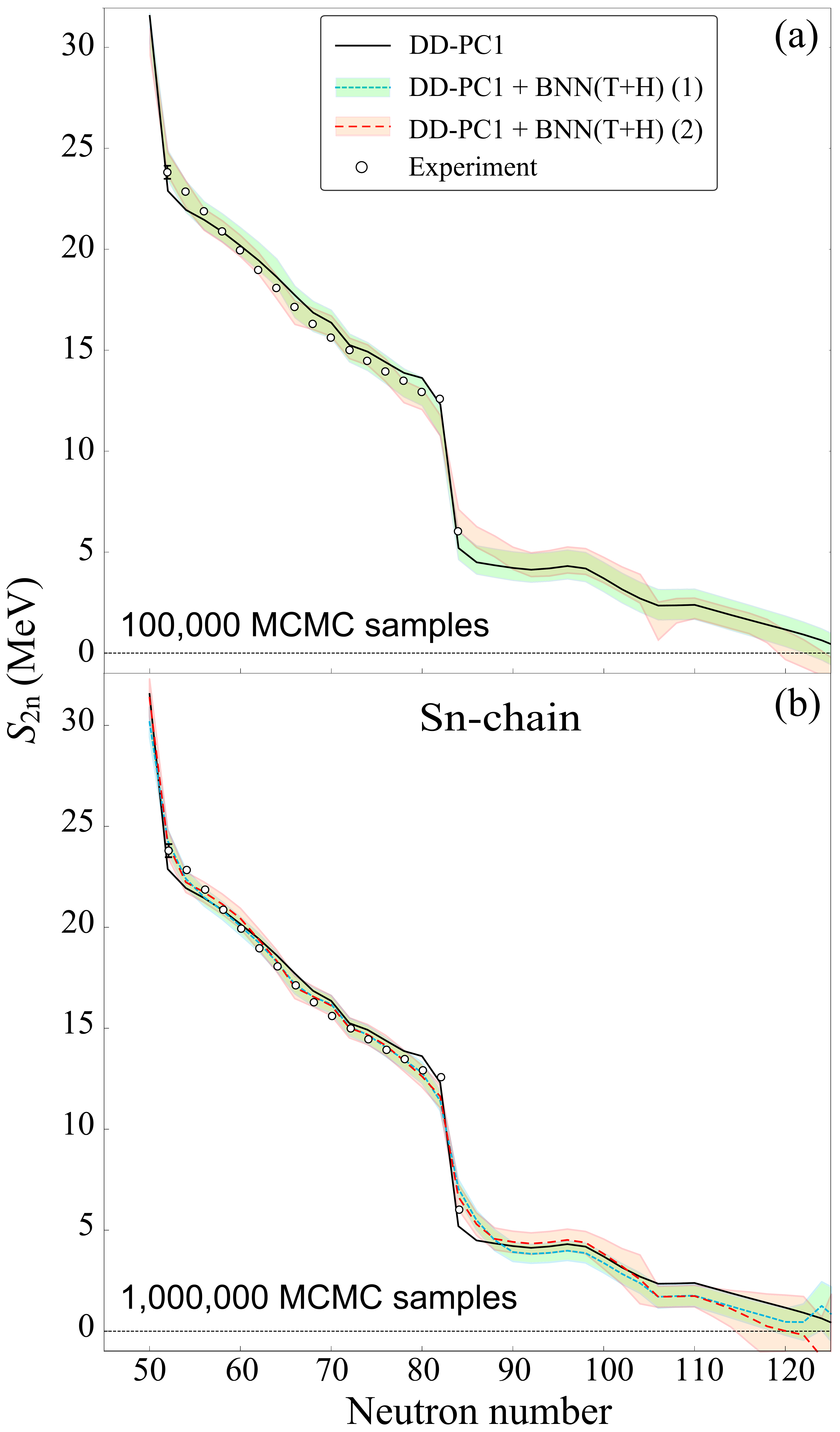}
\caption{Predictions and confidence intervals given by DD-PC1+BNN(T+H) for the Sn chain
for (a) two MCMC runs using 100,000 samples (after 10,000-sample tuning)  and  (b) two MCMC runs using
1,000,000 samples  (after 100,000-sample tuning). The curve (1) in the top panel corresponds to
the BNN(T+H)  curve of Fig.~\ref{extrapolations-sn}(d).}
\label{fig-samples-bnn}
\end{figure}

Despite the apparent strong performance of BNN, it is
necessary to provide some caveats based on our experience. Even when using
up to 10$^7$  Monte Carlo samples, the results are not completely stable
throughout different simulation runs, in particular for extreme
extrapolations. We illustrate this in Fig.~\ref{fig-samples-bnn} where we
superpose the predictions and confidence intervals given by the  DD-PC1 + BNN(T+H) model trained on AME2016+JYFLTRAP dataset for the 
Sn-chain, with two different MCMC runs with 100,000 samples (after 10,000-sample tuning) and 1,000,000 samples  (after 100,000-sample tuning).  Obviously one would expect the curves to match, but they can differ here significantly, even with a high number of iterations. 
%%%%
%%%%%%

\begin{figure}[htb]
\includegraphics[width=0.9\linewidth]{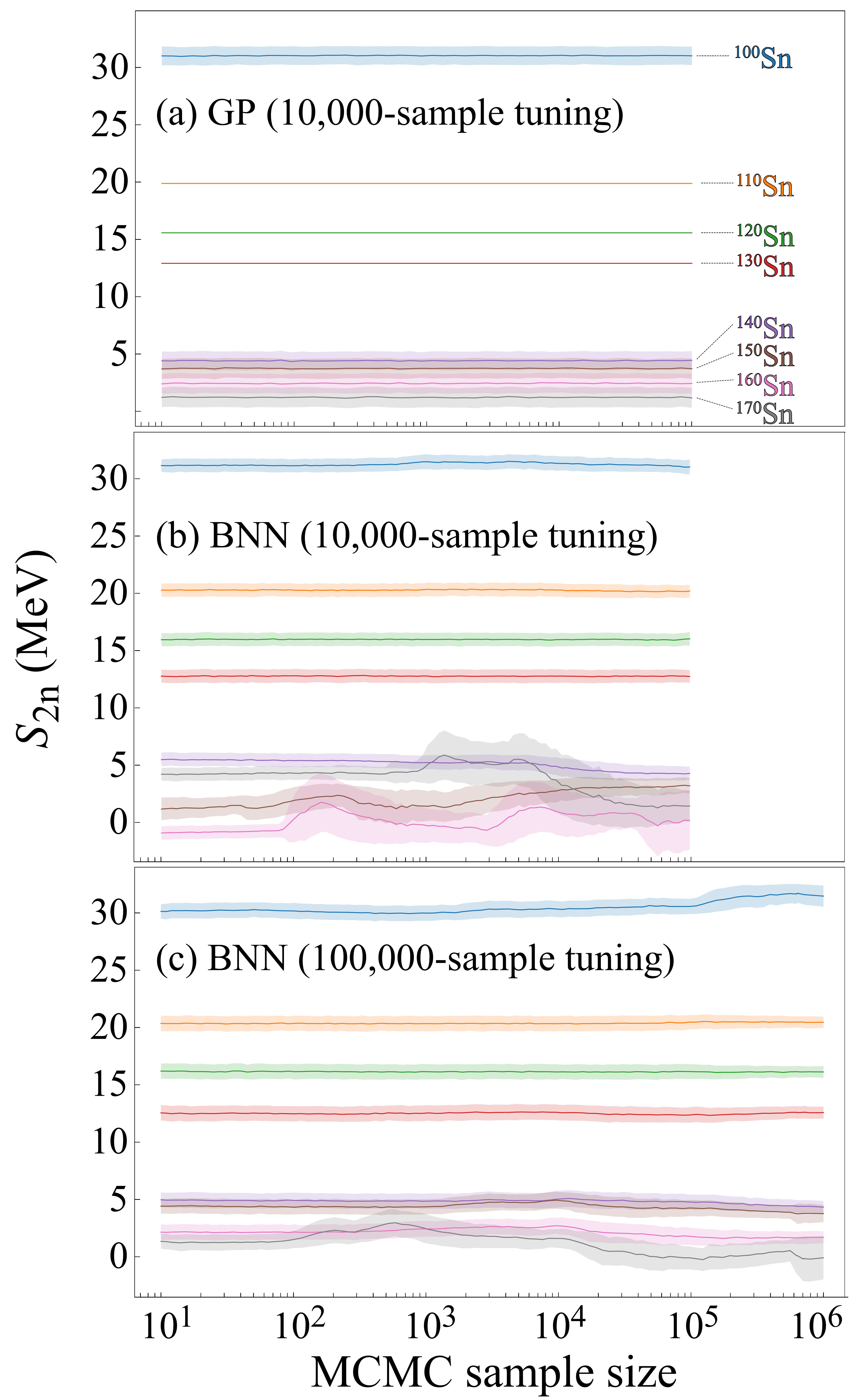}
\caption{Posterior sample mean and standard deviation of $S_{2n}$ for several Sn isotopes
predicted in  DD-PC1. The statistical calculations were carried out with GP(T+H)  and BNN(T+H) methods
 trained on the AME2016+JYFLTRAP  dataset. The results are shown as a function of the
number of   MCMC samples. The number of samples used initially for tuning was
10,000  (panels a and b) and 100,000  (panel c).
}
\label{fig-convergence-combined}
\end{figure}

Strong evidence exists to argue that the
numerical instabilities inherent to BNN are  related to the large number of parameters
used.  Examples of numerical
errors on real data problems, associated with convergence difficulties, can be found  in Ch.\,4.4 of Ref.~\cite{Neal}. A systematic
investigation of discrepancies between several BNN runs is difficult to evaluate
for our problem, because it would require extensive computations.
However, investigating the convergence of BNN can be done at a relatively
low cost. This leads to the results  shown in Fig.~\ref{fig-convergence-combined}. 
%%%%
%%%%%%

One can see
that after using 10,000 samples for tuning, the posterior predictions and
uncertainties of the GP emulator are clearly in a stationary regime, but it
takes about 100,000 samples for BNN to reach the same level.  In general, convergence of MCMC estimators is governed by
the Central Limit Theorem (CLT) according to which the convergence rate behaves as $\frac{C}{\sqrt{n}}$, where the constant $C$ corresponds to the largest
eigenvalue of the autocovariance matrix of the Markov chain \cite{MC-CLT}.
For models with many parameters such as BNN, correlation between samples
cannot be avoided with the traditional algorithms of Metropolis(-Hastings)
or Gibbs. Convergence can be improved by so-called 'variance reduction
techniques' which aim at decreasing the constant $C$. A number of  Bayesian
nuclear mass studies \cite{Utama16,Utama16bis,Utama17,Utama18,Niu2018}
has relied on the   ``Flexible Bayesian
Modeling" software \cite{Neal},  which is based on a combination of Gibbs and
Metropolis algorithms,  which can improve numerical
convergence. Other suggestions in Ref.~\cite{Neal} to improve convergence  include changing
priors to non-Gaussian stable distributions or using simulated annealing.
Modern approaches favor the use of more advanced MCMC techniques relying on
an energy density known as Hamiltonian Monte Carlo methods such as NUTS \cite{NUTS,HMC}, which explore the space more wisely and converge in less steps, however,
at some significant increased cost in terms of evaluation time. The
underlying idea is that convergence of Markov chains on bounded spaces
occurs at the exponential rate $\rho ^{n}$ where $0<\rho \leq 1$ is the
largest eigenvalue of the transition matrix of the parameters. This
exponential convergence can only be achieved when the proposal distribution
matches the actual distribution, which requires an adequate exploration of
the space \cite{MC-exp}. In our simulations, however, NUTS and Metropolis
sampler achieved similar convergence for comparable computation time
although fine tuning could improve NUTS. Because of the CLT limitation, our
best option to improve the convergence of BNN would be to
increase the number of simulations by two orders of magnitude, which would
require unreasonable levels of computational resources, making the method
impractical.

All the numerical issues encountered in BNN are essentially absent in the
case of GP, which has a very small number of parameters (three)
in comparison to BNN (over one hundred). This numerical
stability is an additional argument, beyond parsimony, in favor of GP over
BNN. We must mention that the evaluation of the prediction samples from the
model parameters requires sampling from a multivariate Gaussian kernel,
which can take significantly more time than in the case of BNN, but far less
than the factor of 100 which would be needed to improve BNN's stability.

\section{Conclusions}
\label{Conclusions}

There is a vast amount of information contained in the residuals of theoretical models' predictions.
To improve the fidelity of theoretical predictions, especially in the context of extrapolations, 
one can  utilize Bayesian machine learning techniques such as GP or BNN, which can quantify patterns of deviations between theory and experiment. Stochastic simulations based on MCMC sampling provide statistical corrections to average prediction values, and offer full uncertainty quantification on predictions through credibility intervals. 

In this study, we investigated patterns of $2n$ separation energies of even-even nuclei calculated by several global mass models. We proceeded in three steps. First, we trained our statistical models on large learning datasets of experimentally known $S_{2n}$ values. We then made extrapolative predictions 
for the testing datasets consisting of more recently measured separation energies,
as a way to validate the statistical method's predictive performance. Having thus established and validated a statistical methodology, including a determination of its parameters, we then carried out predictions for unknown data. 

We see that although both GP or BNN reduce the rms deviation from experiment significantly,
GP has a better and more stable performance (see Tables \ref{table_2016} and %
\ref{table_2017}). Both GP and BNN perform best on the relativistic DFT
models, second on the Skyrme-DFT models, and then on the more
phenomenological models FRDM-2012 and HFB-24, which are very well optimized
to nuclear masses; this is consistent with the corresponding levels of
structure that one could expect. The increase in the predictive power of
DFT-based models aided by the statistical treatment is quite astonishing:
the resulting rms deviation from experiment in the T+H variant (last column
of Table~\ref{table_2016}) shows that the carefully optimized Skyrme-DFT
models, such as UNEDF0, UNEDF1, and SV-min, provide results of a similar
quality on the testing dataset as the more phenomenological models.
Overall, for the testing dataset AME2016-2003, the rms deviation from
experimental $S_{2n}$ values is in the 400-500\,keV range in the GP
T+H variant for the \textit{all} theoretical models employed in this study.

We realized that, as the classical sigmoid activation function used in BNN has
linear tails that do not vanish, it is poorly suited for a bounded
extrapolation. This perhaps explains the better performance of GP, which
builds the prediction locally with small bumps that can capture local
trends. Indeed, our results indicate that the $S_{2n}$-residuals have an appreciable
local structure around magic gaps, but no global structure. This is encouraging as it means
that the models used are not missing any significant physics applicable on
the whole nuclear domain. 
As a broader perspective, our results  support using local Bayesian
statistical models in combination with a well thought-out effect range, to
reproduce the residuals trends, and to take advantage of models and other
forms of physical intuition as part of building a Bayesian prior.
At the same time, BNN should not be rejected as a poor extrapolation tool at all ranges. We emphasize that the statistical corrections and quantified uncertainties obtained by GP on extrapolations far from the range of the training data are negligible in practice, which is by design of the GP specification. It is also true that BNN extrapolations are possible beyond the range of influence of GP. However, in the absence of any supporting experimental data to test the performance of BNN far from the stability range, it is not possible to know whether the actual BNN corrections are of value in these long-range extrapolations. We contend that the main interest in Bayesian methods when applied to distant extrapolations lies in the UQ that their credibility intervals provide. As soon as a few data points in these distant ranges will become available, it will be possible to test BNN's extrapolation performance using UQ as a framework for honest performance metrics.

Our Bayesian methodology is very robust in the sense of this type of performance framework. To this point, we showed how the ECP curves we obtain
match the reference values very well, in a slightly conservative way in most
cases, which is highly desirable to ensure UQ honesty without being wasteful
(cf. discussion around Fig.~\ref{ecp}).

The statistical approach to extrapolation of nuclear model results discussed
in this paper can be very useful for assessing the impact of current and
future experiments  in the context of model developments. In the particular
case studied in this work, the impact of the new data for unstable nuclei on
the predictive power of theoretical models turned out to be minor. This is
probably due the fact that the mass surface in the region of JYFLTRAP data
is fairly smooth. This conclusion should not be generalized; rather, because
the methodology provide a sound UQ, for instance, one should expect that any
experiment planned near the boundary of a mass surface with limited
smoothness should provide a significant advantage to extrapolations beyond
that boundary. Other scenarios can also be imagined with similar positive
impact of extrapolation from a small number of new data points near a
boundary region. Such a scenario can be tested ahead of time, using
synthetic data, with our methodology providing a full quantification of
uncertainty based on the synthetic scenarii, to help experimenters decide if
the experiment's cost is worth the risk.

We also think that the new GP capability to evaluate residuals is expected
to impact research in the domains where experiments are currently
impossible, e.g., in studies of astrophysical nucleosynthesis processes. 
%%%%
\begin{figure}[htb]
\includegraphics[width=1.0\linewidth]{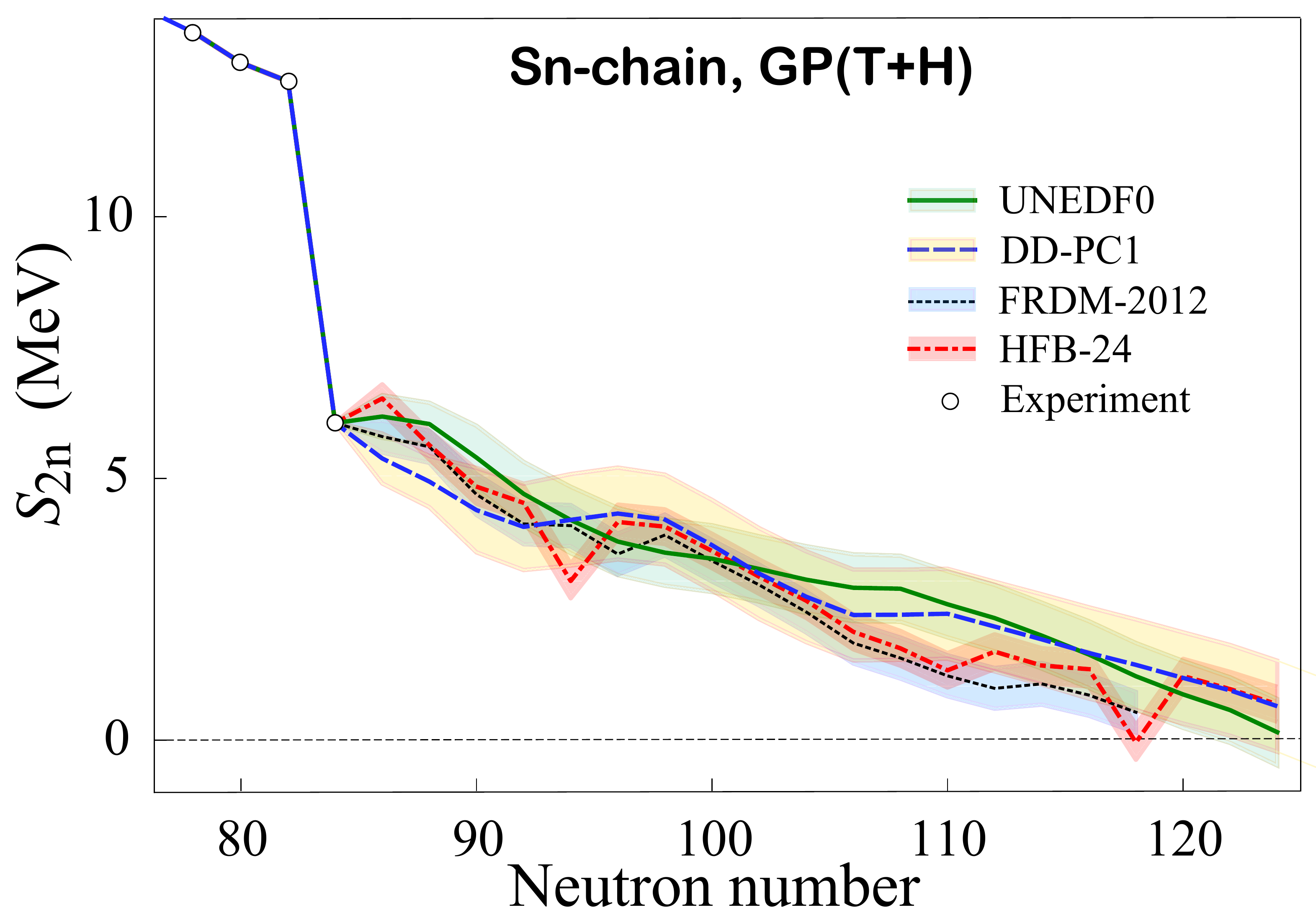}
\caption{Extrapolations of $S_{2n}(Z,N)$ for the Sn chain corrected with
GP(T+H) and one-sigma CIs, combined for four
representative models. The different
models are consistent overall once the statistical correction and uncertainty are taken into  account.}
\label{fig-extrapolations-gp-combined}
\end{figure}
%%%%%%
Figure~\ref{fig-extrapolations-gp-combined} illustrates this point nicely. It shows  extrapolations of the $S_{2n}$-values for the even-even Sn isotopes for four global mass models aided by the statistical GP(T+H) approach. It is gratifying to see that all four models displayed provide internally consistent results, i.e., they agree with each other within the estimated CIs.

The example shown in Fig.~\ref{fig-extrapolations-gp-combined} illustrates the limited value of assessing physics models solely based on their ability to fit experimental data.
Clearly, mean values  tell only part of the story. While HFB-24 and FRDM-2012 provide the superior reproduction of measured masses, their ability to extrapolate into the $r$-process region of neutron-rich nuclei is similar to that of DFT models based on  well-optimized energy density functionals.  
Going further, we can strengthen our statements and
decrease the sizes of uncertainty regions by combining the predictions'
CIs provided by GP and BNN for one mass model,
and even of several mass models with Bayesian corrections (see Fig.~\ref{fig-extrapolations-gp-combined}). For a na\"{\i}ve look at this idea,  one  may consider the intersection of the various CIs.

When it comes to astrophysical applications, such as simulations of nucleosynthesis and models of the composition and structure of the neutron star crust, the methodology presented in this work can be directly applied to each observable of interest, such as one-neutron separation energies and $Q$-values for beta decay, alpha decay, and fission. The strategy, which we are going to adopt, is to provide Bayesian corrections and covariances for theoretical mass tables.  Such information will allow the statistical determination of all mass differences, including precise statements about estimation precision, via a full quantification of their uncertainties.

There are other ways of further reducing theoretical uncertainty. For
instance, it may be possible to decrease the residuals locally by
fine-tuning model parameters to selected regional data. In this respect,
measurements of masses of more exotic nuclei at rare isotope facilities will
greatly add to the dataset that can be used in such analyses. 
Another way is to combine the predictions of several statistically corrected
nuclear models, such as these shown in Fig.~\ref{fig-extrapolations-gp-combined}.
Two Bayesian approaches can be used \cite{Hoeting1999,Was00} in this context: 
model selection (the problem of using the data to select one model from a list of candidate models) and model averaging (estimating some quantity under each model and then averaging the estimates according to how likely each model is).
In particular, an extrapolation based on a given
model can run a significant misspecification risk, and this risk is
typically never taken into account. Under Bayesian model selection/averaging, this risk can be quantified
and taken into account, always resulting in a more honest UQ, and sometimes
in more accurate extrapolation. These developments are left for a future
study.

Finally, let us note that  the information contained in the residuals shown, e.g.,  in Figs.~\ref{raw-residuals} and \ref{smoothed-residuals}, provides crucial guidance for the further developments of nuclear mass models. While statistical methods, which couple current nuclear models with available experimental data to maximize the use of existing information,  can help providing more reliable predictions, they cannot be a substitute for the systematic development
of high-fidelity theoretical models of the nucleus. The hope is that such models, guided by statistical methods, will allow us uncover nuclear structure features that may appear only far from stability.

% ACKNOWLEDGMENTs
\begin{acknowledgments} 
Discussions with Jorge Piekarewicz are gratefully appreciated. This work was supported by the U.S. Department of Energy under
Award Numbers DOE-DE-NA0002574 (NNSA, the Stewardship 
Science Academic Alliances program),  and DE-SC0018083 and DE-SC0013365 (Office of Science,  Office of Nuclear Physics). 
\end{acknowledgments}

% BIBLIOGRAPHY
%\bibliographystyle{apsrev4-1}
%\bibliography{Bayes-masses}


\end{document}